\documentclass[apj]{emulateapj}
\usepackage{natbib}
\usepackage{color}
\usepackage[markup=bfit]{changes}
\shorttitle{Nature of MWC\,728}
\shortauthors{A.S. Miroshnichenko et al.}

\begin{document}

\title{Toward Understanding The B[e] Phenomenon: V. Nature and Spectral Variations of the MWC\,728 Binary System.}
\thanks{This paper is partly based on observations obtained
at the Canada-France-Hawaii Telescope (CFHT) which is operated by
the National Research Council of Canada, the Institut National des
Sciences de l$^{\prime}$Univers of the Centre National de la
Recherche Scientifique de France, and the University of Hawaii as
well as on observations obtained at the 2.7\,m Harlan J. Smith
telescope of the McDonald Observatory (Texas, USA), 2.1\,m of the
Observatorio Astronomico Nacional San Pedro Martir (Baja California, Mexico), 2\,m
telescope of the Ond\v{r}ejov Observatory, Czech Republic, and
0.81\,m telescope of the Three College Observatory, North Carolina,
USA.}

\author{A.~S.~Miroshnichenko$^{1,2}$}
\affil{$^1$Department of Physics and Astronomy, University of North
Carolina at Greensboro, Greensboro, NC 27402--6170, USA}
\affil{$^2$ National center of space exploration and technologies, Almaty, Kazakhstan}

\author{S.~V.~Zharikov$^3$}
\affil{$^3$Instituto de Astronom\'ia, Universidad Nacional Aut\'onoma
de Mexico,  Ensenada,  Baja California, 22800, Mexico}

\author{S.~Danford$^1$}

\affil{$^1$Department of Physics and Astronomy, University of North
Carolina at Greensboro, Greensboro, NC 27402--6170, USA}

\author{N.~Manset$^4$}
\affil{$^4$CFHT Corporation, 65--1238 Mamalahoa Hwy, Kamuela, HI
96743, USA}

\author{D.~Kor\v{c}\'akov\'a$^5$, R.~K\v{r}\'{\i}\v{c}ek$^5$}
\affil{$^5$Astronomical Institute, Charles University in Prague, V
Hole\v{s}ovi\v{c}k\'ach 2, 18000, Praha 8, Czech Republic}

\author{M.~\v{S}lechta$^6$}
\affil{$^6$Astronomical Institute of the Academy of Science of the
Czech Republic, Fri\v{c}ova 298, 25165, Ond\v{r}ejov, Czech
Republic}

\author{Ch.~T.~Omarov$^7$, A.~V.~Kusakin$^{2,7}$}
\affil{$^7$Fessenkov Astrophysical Institute, Observatory, 23, Almaty 050020, Kazakhstan}

\author{K.~S.~Kuratov$^{2,7,8}$}
\affil{$^8$ Physico-Technical Department, Al Farabi Kazakh National University, Al Farabi Av., 71, Almaty 050038, Kazakhstan}

\author{K.~N.~Grankin$^9$}
\affil{$^9$Crimean Astrophysical Observatory, Scientific Research
institute, 298409, Nauchny, Crimea, Russia}

\begin{abstract}
We report the results of a long-term spectroscopic monitoring of the
FS\,CMa type object MWC\,728. We found that it is a binary system
with a B5 {\sc V}e (T$_{\rm eff}$ = 14000$\pm$1000 K) primary and a
G8 {\sc III} type (T$_{\rm eff} \sim$ 5000 K) secondary. Absorption line positions of
the secondary vary with a semi-amplitude of $\sim$20 km\,s$^{-1}$ and
a period of 27.5 days. The system's mass function is
2.3$\times10^{-2}$ M$_\odot$, and its orbital plane is
$\sim$13--15$\arcdeg$ tilted from the plane of the sky.
The primary's $v \sin i \sim$110 km\,s$^{-1}$ combined with this tilt
implies that it rotates at a nearly breakup velocity.
We detected strong variations of the Balmer and He {\sc i} emission-line
profiles on timescales from days to years. This points to a variable
stellar wind of the primary in addition to the presence of a
circum-primary gaseous disk. The strength of the absorption-line
spectrum along with the optical and near-IR continuum suggest that
the primary contributes $\sim$60\% of the $V$--band flux, the disk
contributes $\sim$30\%, and the secondary $\sim$10\%.
The system parameters, along with the interstellar extinction, suggest a
distance of $\sim$1 kpc, that the secondary does not fill its Roche lobe,
and that the companions' mass ratio is $q \sim$0.5.
Overall, the observed spectral variability and the presence of
a strong IR-excess are in agreement with a model of a close binary
system that has undergone a non-conservative mass-transfer.
\end{abstract}


\section{Introduction} \label{intro}

This paper continues a series devoted to studying objects from the
Galactic FS\,CMa type group consisting of $\sim$70 members and
candidates characterized by the B[e] phenomenon. The phenomenon
was discovered by \citet{as76} and refers to the presence of forbidden emission lines of [Fe
{\sc ii}]  and/or [O  {\sc i}] in the spectra of B--type stars. In
the vast majority of cases one also detects a strong excess of
IR radiation that is attributed to circumstellar dust. \citet{l98}
recognized four subgroups of B[e] objects with known evolutionary
status: pre-main-sequence Herbig Ae/Be stars (``HAeB[e]''),
symbiotic binaries (``SymB[e]''), compact Planetary Nebulae
(``cPNB[e]''), and supergiants (``sgB[e]''). They concluded that the
B[e] phenomenon is found in objects
at very different evolutionary stages but with similar conditions
in their circumstellar envelopes. However, they were unable to
classify $\sim$50\% of the objects selected by \citet{as76} and
called them unclassified objects with the B[e] phenomenon (``unclB[e]'').

\begin{figure}[t]
\begin{center}
\includegraphics[width=8cm]{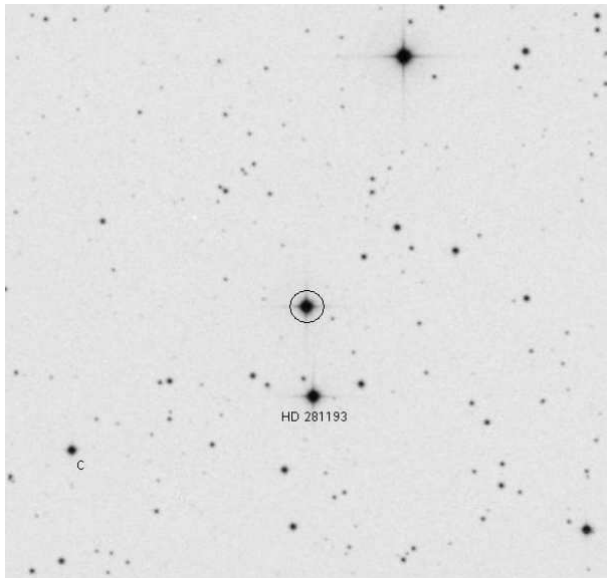}
\end{center}
\caption{{A $10\arcmin \times 10\arcmin$ field around MWC\,728
(marked by the circle). The comparison star (HD\,281193) and check
star (NOMAD\,1197$-$46760 labeled with a ``C'') are shown.  North is
at the top and East is at the left.} \label{f:field}}
\end{figure}

\begin{table}[b]
\caption[]{Summary of the spectroscopic observations}\label{t1}
\begin{center}
\begin{tabular}{lclcr}
\hline\noalign{\smallskip}
Observatory   & Dates  &  Range  & Resolution  & No.\\
\noalign{\smallskip}\hline\noalign{\smallskip}
CFHT            & 2004--2010 &   3600--10500  &   65000  &   3  \\
OAN--SPM    & 2005--2014 &   3600--8000    &   18000  &  36  \\
McDonald      & 2005--2013 &   3600--10500  &   60000  &  11  \\
TCO              & 2013--2014 &   4250--7850    &   10000  &  56  \\
Ond\v{r}ejov  & 2012--2014 &   6250--6750    &   12500  &  10  \\
\noalign{\smallskip}\hline
\end{tabular}
\end{center}
\begin{list}{}
\item Column information: (1) -- Observatory name, (2) -- range of observing dates,
(3) -- spectral range in \AA, (4) - spectral resolving power, and
(5) -- numbers of spectra obtained.
\item Spectrographs used: ESPaDOnS \citep{donati97} at CFHT, REOSC at Observatorio Astronomico Nacional San Pedro Martir
(OAN-SPM), cs23 \citep{Tull95} at the McDonald Observatory, Eshel
from Shelyak Instruments at the Three College Observatory (TCO), and
a Coud\'e spectrograph \citep{Slechta2002} at Ond\v{r}ejov.
\end{list}
\end{table}

\begin{figure}[!t]
\setlength{\unitlength}{1mm}
\resizebox{15.cm}{!}{
\begin{picture}(150,80)(0,0)
\put (-10,0) {{\includegraphics[width=10.7cm]{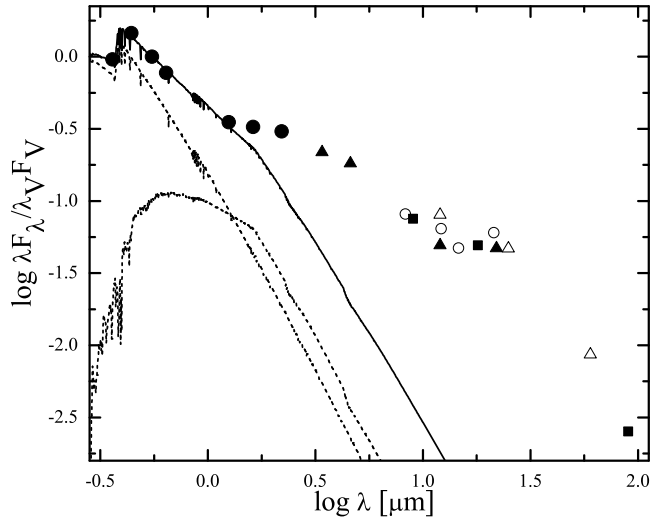}} }

\end{picture}}
\caption{The spectral energy distribution of MWC\,728. Logarithm of
the flux normalized to that in the $V$--band is plotted
vs. logarithm of the wavelength in microns. Symbols show the following data sets:
{\it circles} -- optical and near-IR photometry from Table\,\ref{t2}, {\it
upward filled triangles} -- WISE data \citep{Wright2010}, {\it
upward open triangles} -- IRAS data, and filled squares -- AKARI
data \citep{Murakami2007}. The fluxes were dereddened using the
interstellar extension law from \cite{Savage:1979aa}.  The dashed
lines represent \cite{Kurucz:1994aa} model atmospheres for the hot
star with T$_{\rm eff}$=14000 K and the cool star with
T$_{\rm eff}$=5000 K. The solid line shows total fluxes with a 60\%
contribution from the hot companion, 10\% from the
cool companion, and 30\% from the circum-primary gaseous disk
to the $V$--band flux. \label{f:spectrum}}
\end{figure}

The unclB[e] objects were critically reconsidered by \citet{m07}.
Several possible evolutionary states (e.g., pre-main-sequence stars
or supergiants) were rejected, and the group was renamed FS\,CMa
objects. The main observational properties of this group include the
following: 1) early--B to early--A type optical continuum with
strong emission lines of hydrogen, Fe {\sc ii}, [O {\sc i}], and
sometimes of [Fe {\sc ii}] and [O {\sc iii}] (some absorption lines
from the hot star atmosphere may be present as well, but they are
frequently veiled by the circumstellar continuum); 2) a large IR
excess that peaks at 10--30 $\mu$m and sharply decreases longward;
3) location outside of star-forming regions; and 4) a secondary
companion (so far discovered in about one third of the group
members) which can be a fainter and cooler normal star or a
degenerate object.

\citet{m07a} expanded the group with 10 newly discovered objects found in the
{\it IRAS} database by cross-identification with catalogs of optical
positions. Later \citet{m11} reported $\sim$20 more candidates found
in the NOMAD catalog \citep{z05} using optical and near-IR color
criteria. Although the origin of the B[e] phenomenon is understood
at least qualitatively in all the groups with known evolutionary
status, it still remains controversial whether FS\,CMa objects are close
binaries recently evolved off main-sequence or a distinct group of proto-planetary
nebulae \citep[cf.][]{m07}.

The subject of the present paper is a relatively bright ($V\sim10$ mag)
emission-line star MWC\,728
that was first reported by \citet{mb49}. Some other common
identifications of the object are IRAS\,03421+2935, HD\,281192, and
BD+29$^{\circ}$611. A brief survey of observations of MWC\,728
found in the literature is given by \citet{m07a}, who described
its optical and near-IR spectrum and included it in the
FS CMa objects group. They reported the following spectral features:
double-peaked Balmer and [O~{\sc i}] lines in emission, He {\sc
i} 4471 \AA \ and Mg {\sc ii} 4481 \AA \ absorption line from a
mid--B spectral type star, and absorption lines of  Fe {\sc i}
5328 \AA, Li {\sc i} 6708 \AA, and Ca {\sc i} 6717 \AA\ lines which
indicated the presence of a cool star. We have
been monitoring MWC\,728 for about 10 years using high-resolution
optical spectroscopy and multicolor photometry
to study its nature and evolutionary state.

Details of the observations are provided in Sect.
\ref{observations}. Analysis of the optical spectra and behavior of
the absorption and emission lines is presented in Sect.
\ref{analysis}. Suggestions about the system's nature are discussed
in Sect. \ref{discussion}, and conclusions are summarized in Sect.
\ref{conclusions}.

\section{Observations}\label{observations}

Optical spectroscopic observations of MWC\,728 were obtained between
December 2004 and February 2015. A summary of the facilities and
instruments used in this project is given in Table \ref{t1}. We have
obtained 116 spectra with a range of spectral resolving powers
of $R=10000-65000$ and signal-to-noise ratios from $\sim$30 to
$\ge$200. The spectra obtained at the McDonald Observatory (Texas,
USA), Observatorio Astronomico Nacional San Pedro Martir (OAN SPM,
Baja California, Mexico), Ond\v{r}ejov Observatory (Czech Republic),
and Three College Observatory (TCO, North Carolina, USA) were
reduced in a standard way with the {\it echelle/slit} package in
IRAF\footnote{IRAF is distributed by the National Optical Astronomy
Observatory, which is operated by the Association of Universities
for Research in Astronomy (AURA) under a cooperative agreement with
the National Science Foundation.}. Observations obtained at the CFHT
were reduced with the Upena and Libre-ESpRIT \citep{donati97}
software packages. Typical uncertainties of the wavelength
calibration are $<$1\,km\,s$^{-1}$ for the CFHT and McDonald data
and $\sim$1--2\,km\,s$^{-1}$ for the other observatories data.
Individual radial velocity measurements may have larger errors
depending on the line strength and the profile shape.

\begin{table}[ht]
\caption[]{Photometric observations of MWC\,728} \label{t2}
\begin{center}
\begin{tabular}{ccrccccc}
\hline\noalign{\smallskip}
JD &  $V$ &  $U-B$  & $B-V$& $V-R$& $J$  & $H$  & $K$  \\
 2450000+ &  &  &&&&& \\
\noalign{\smallskip}\hline\noalign{\smallskip}
4748.3885   & 9.79 &  0.16   & 0.32 & 0.31 &      &      &      \\
4750.4548   & 9.75 &  0.04   & 0.23 & 0.31 &      &      &      \\
4753.5535   &      &         &      &      & 8.28 & 7.46 & 6.90 \\
4755.5736   &      &         &      &      & 8.49 & 7.62 & 6.74 \\
5140.9304   &      &         &      &      & 8.32 & 7.45 & 6.76 \\
5145.5718   & 9.80 & $-$0.17 & 0.25 & 0.40 &      &      &      \\
5487.9896   &      &         &      &      & 8.32 & 7.45 & 6.66 \\
5488.9299   &      &         &      &      & 8.38 & 7.51 & 6.73 \\
\noalign{\smallskip}\hline
\end{tabular}
\end{center}
\vspace*{-0.3cm}
\begin{list}{}
\item The $UBVR_{\rm J}$ observations were obtained at CrAO and the $JHK$ observations
were obtained at OAN SPM.
\end{list}

\vspace*{0.3cm}
\begin{center}
\begin{tabular}{cccc}
\hline\noalign{\smallskip}
JD 2450000+ &  $V$ &  $B-V$  & $V-R$\\
\hline\noalign{\smallskip}
6917.265    &   9.80    &   0.31    &   0.22    \\
6918.269    &   9.85    &   0.29    &   0.24    \\
6945.429    &   9.76    &   0.27    &   0.23    \\
6948.458    &   9.78    &   0.28    &   0.24    \\
6960.396    &   9.85    &   0.31    &   0.26    \\
6963.485    &   9.78    &   0.31    &   0.27    \\
6972.184    &   9.78    &   0.27    &   0.23    \\
6984.312    &   9.79    &   0.27    &   0.24    \\
6988.188    &   9.79    &   0.26    &   0.24    \\
6993.251    &   9.77    &   0.28    &   0.23    \\
7000.182    &   9.75    &   0.28    &   0.22    \\
7001.170    &   9.77    &   0.28    &   0.22    \\
7002.262    &   9.75    &   0.27    &   0.22    \\
7003.297    &   9.77    &   0.27    &   0.21    \\
7005.326    &   9.80    &   0.25    &   0.24    \\
7006.295    &   9.75    &   0.27    &   0.22    \\
7019.167    &   9.78    &   0.27    &   0.22    \\
7020.149    &   9.77    &   0.29    &   0.23    \\
7022.133    &   9.81    &   0.28    &   0.24    \\
7023.242    &   9.79    &   0.28    &   0.24    \\
7050.266    &   9.75    &   0.28    &   0.22    \\
7058.242    &   9.89    &   0.33    &   0.24    \\
7059.054    &  10.01   &   0.37    &   0.28    \\
\noalign{\smallskip}\hline
\end{tabular}
\end{center}
\vspace*{-0.3cm}
\begin{list}{}
\item The $BVR_{\rm C}$ observations obtained at TShAO in the
Johnson-Cousins photometric system. \end{list}
\end{table}

Photometric data for MWC\,728 were obtained at the following sites.
Three $UBVR_{\rm J}$ observations were obtained at a 0.6\,m
telescope of the Crimean Astrophysical Observatory (CrAO) with a
single-element photometer in 2008 and 2009 \citep[for the data
reduction technique see][]{gr2008}. $BVR_{\rm C}$ observations
of a 10$^{\prime}\times10^{\prime}$ field around the object were
obtained on 23 nights at a 1\,m telescope of the Tien-Shan Observatory (TSAO) of
the Fesenkov Astrophysical Institute of the National Academy of
Sciences of Kazakhstan between September 2014 and February 2015.
A 3056$\times$3056 Apogee F9000 D9 CCD camera with 12$\mu$m pixels
and a 3--step cooling was used with a set of $BVR$ filters. The TSAO
data were reduced using Maxim-DL, while brightness of the object and
field stars were measured using the IRAF task {\it imexamine}. The TSAO instrumental
magnitudes were converted into the standard Johnson-Cousins $BVR$ system using data
for several open clusters.
The TSAO photometry was calibrated by comparing the brightness of MWC\,728 to
HD\,281193 as a comparison star and NOMAD\,1197$-$46760 as
a check star (see Figure\,\ref{f:field}). The brightness difference between the comparison and check
star was found to be $\le$ 0.01 mag in all our images. Photometry of HD\,281193
was obtained at CrAO and resulted in the following magnitudes: $V$ =
10.00 mag, $B-V$ = 0.28 mag, $V-R_{\rm J}$ = 0.18 mag.

Near-IR $JHK$ observations were obtained at the 0.84\,m telescope of
the OAN SPM on 5 nights in 2008--2010 with a 256$\times$256 pixels
IR camera Camila \citep{gg94}. The images were calibrated by
observing standard stars from \citet{hunt98}. The near-IR brightness of MWC\,728
in our data is $\sim$20\% higher than that in the 2MASS catalog recorded in
December 1999 \citep[cf.][]{m07a}. The results for all the described photometric
observations are presented in Table \ref{t2}.  The accuracy of all the data is 0.02--0.03 mag.

\begin{figure*}[!t]
\setlength{\unitlength}{1mm}
\resizebox{15.cm}{!}{
\begin{picture}(150,70)(0,0)
\put (0,0) {{\includegraphics[width=8.75cm, bb = 45 40 720 522, clip=]{MWC728_fig3a.eps}} }
\put (95,0) {{\includegraphics[width=8.25cm,bb = 10 10 300 260, clip=]{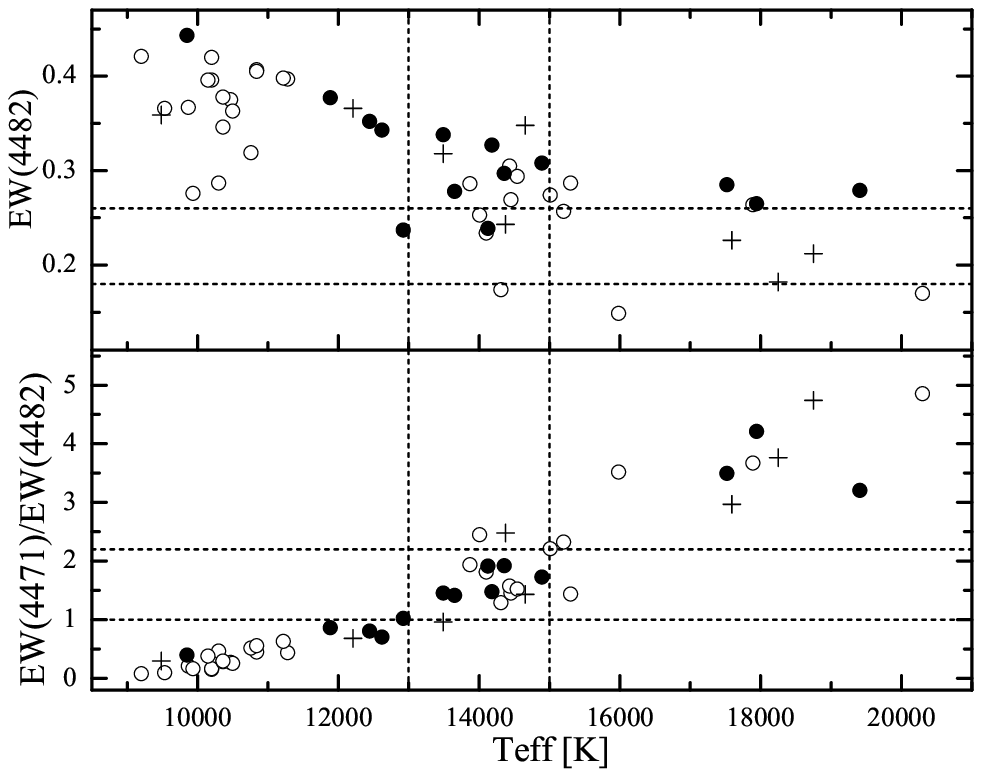}}}
\end{picture}}
\caption{{\bf Left panel:} Comparison of the He {\sc i} 4471 \AA\
and the Mg {\sc ii} 4482 \AA\ lines in the spectrum of MWC\,728
taken on 2008 December 15 at McDonald with those of normal stars.
The spectrum of BS\,4967 (B5/6 {\sc iii}, $v \sin i$ = 115
km\,s$^{-1}$) was taken at TCO, and the spectrum of BS\,896 (B6 {\sc
iii}, $v \sin i$ = 105 km\,s$^{-1}$) was taken at the OHP. Intensity
is normalized to the nearby continuum, while the spectra were moved
to the same He {\sc i} line position for all the stars.
{\bf Right panel:} T$_{\rm eff}$ relationship with EWs in \AA\ of the
He {\sc i} 4471 \AA\ and the Mg {\sc ii} 4482 \AA\ lines. Circles
represent data for normal B--type stars (filled -- OHP, open -- TCO)
and crosses represent OHP data for Be stars. Temperature
determinations were collected from various papers
\citep[e.g.,][]{Fremat05,Zorec09,Zorec12}. The horizontal dashed lines show
ranges of the EW variations detected in the spectra of MWC\,728, and
the vertical dashed lines show most likely boundaries of the T$_{\rm eff}$ for
MWC\,728 (see explanations in Sect.\,\ref{absorptions}). \label{f:HeMgTeff}}
\end{figure*}

\begin{figure}[!t]
\setlength{\unitlength}{1mm}
\resizebox{15.cm}{!}{
\begin{picture}(150,80)(0,0)
\put (0,0) {{\includegraphics[width=8.75cm,bb = 0 0 270 230, clip=]{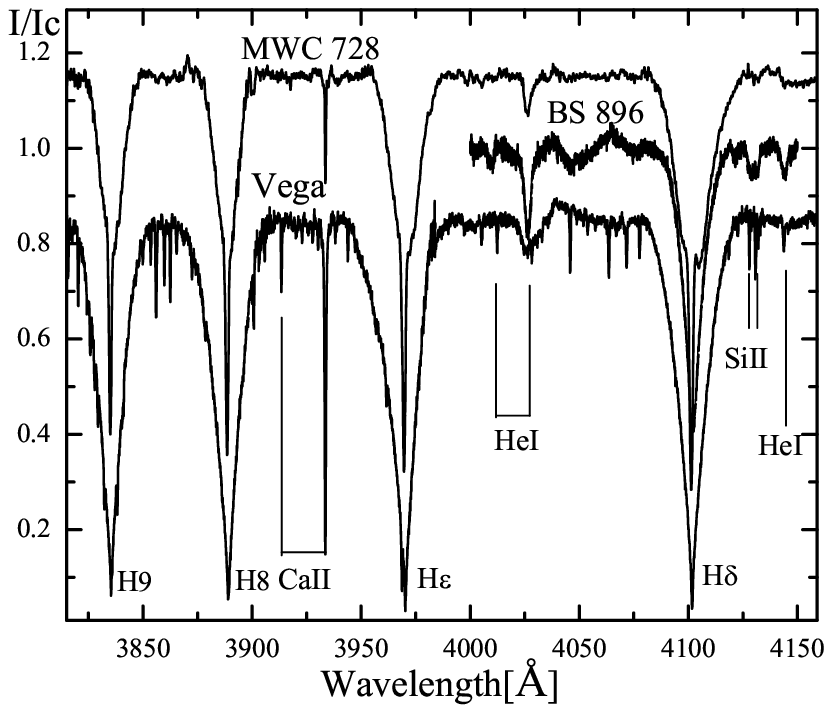}}}
\end{picture}}
\caption{Comparison of the blue part of the spectrum of MWC\,728
with those of BS\,896 (B6 {\sc iii}) and Vega (A0 {\sc v}). The
spectra of Vega and MWC\,728 were obtained at OAN SPM, while the
spectrum of BS\,896 was taken at OHP (4000 \AA\ is its shortest
wavelength). \label{f:HighBalmer}}
\end{figure}

\section{Data Analysis}\label{analysis}

The spectral energy distribution of MWC\,728 is shown in Figure~\ref{f:spectrum}.
It is characterized by a slightly reddened optical continuum of a hot star and
a strong IR excess produced by circumstellar gas and dust.
A detailed analysis of the optical absorption
and emission lines is presented in the following subsections, while the components
of the total flux are discussed in Sect.\,\ref{discussion}. The dust component of the
system will be analyzed in a follow-up paper.

\subsection{Absorption lines}\label{absorptions}

Absorption lines in the spectrum of MWC\,728 can be divided into
three groups.

The first group contains relatively broad (FWHM $\sim$220
km\,s$^{-1}$) and weak (I/I$_{\rm cont} \le$ 0.9) lines that
manifest a moderate projected rotation of a hot star (see Figure\,\ref{f:HeMgTeff}, left panel).
It includes He {\sc i} lines (4009, 4026, 4471, 5876, 6678 \AA), the Mg
{\sc ii} 4482 \AA, Si {\sc ii} 6347 and 6371 \AA, and the O {\sc i}
7772--7775 \AA\ triplet.

The Mg {\sc ii} 4482 \AA\ line has an average equivalent width (EW) of
0.22$\pm$0.03 \AA\ which is typically 1.7$\pm$0.3 times weaker than
the EW of the He {\sc i} 4471 \AA\ line. The EW ratio of these lines
serves as an effective temperature (T$_{\rm eff}$) criterion. We used
spectra of $\sim$60 B--type dwarfs and giants as well as Be stars
obtained at the Observatoire de Haute-Provence
\citep[OHP, $R \sim$48000,][]{moul04} and TCO to calibrate the ratio
(Figure\,\ref{f:HeMgTeff}, right panel). Since the EW of both lines can be
affected by a contribution from the cool star and the circumstellar material,
we need to explore this effect.

As shown below in this Section, the cool star makes a negligible
contribution in this spectral region.
Any additional circumstellar conti\-nuum would lower EWs of both
lines. Therefore, removal of such a continuum would lead to a lower
temperature estimate based on the Mg {\sc ii} EW but to a higher one
based on the He {\sc i} EW. Although the lines may also be affected
by radiation transfer in the disk, the net effect is probably small
judging from the positions of Be stars in the right panel of
Figure\,\ref{f:HeMgTeff}. The He {\sc i} lines in the blue part of the
spectrum of MWC\,728 (Figure\,\ref{f:HighBalmer}) look weaker than
those of BS\,896 (T$_{\rm eff} =$ 14000 K) and comparable to
those of Vega (T$_{\rm eff} =$ 9500 K). However, the circumstellar
contribution to the observed optical conti\-nu\-um estimated in
Sect.\,\ref{discussion} partially accounts for the He {\sc i} lines
weakening. With this analysis, we estimate the hot companion T$_{\rm
eff} = 14000\pm$1000 K which corresponds to a
spectral type B5 \citep{Gray1994}. Additionally, comparison with spectra
of B--type stars, which show no evidence for a noticeable amount of
circumstellar gas, suggests a projected rotational velocity of
$v \sin i \sim$110 km\,s$^{-1}$ (see Figure\,\ref{f:HeMgTeff}, left panel).

Atmospheric absorption wings are only occasionally seen in H$\alpha$,
while higher members of the Balmer series do not fill their atmospheric
profiles with emission. The latter exhibit deep and narrow line cores
during periods of the enhanced emission (starting from H$\delta$, see
Fi\-gure\,\ref{f:HighBalmer}). The Si {\sc ii} lines are weak that is typical
of dwarfs and giants, thus excluding a supergiant luminosity.
The Paschen series lines and Fe {\sc ii} lines are seen neither in absorption
nor in emission. They are most likely filled with emission components
as well as veiled by the continuum of the cool companion and circum-primary disk.

\begin{figure}[!t]
\setlength{\unitlength}{1mm}
\resizebox{15.cm}{!}{
\begin{picture}(150,70)(0,0)
\put (0,0) {{\includegraphics[width=8.75cm,bb = 45 40 720 522, clip=]{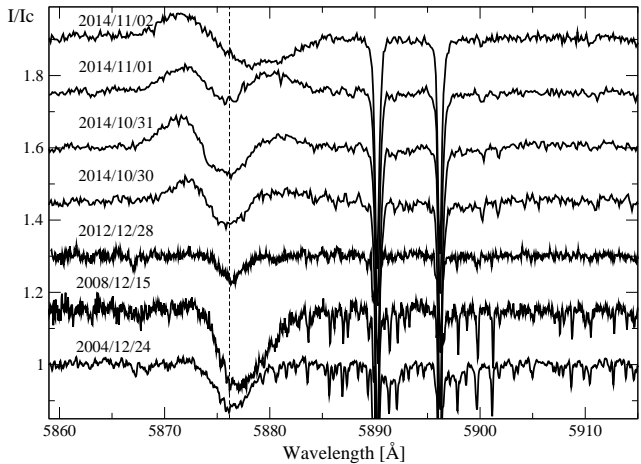}}}
\end{picture}}
\caption{Variations of the He {\sc i} 5876 \AA\ line. The spectra were obtained (from bottom to top) at: CFHT
in 2004, the McDonald Observatory in 2008 and 2012, and OAN SPM in 2014.
The wavelength scale is heliocentric. The vertical dashed line shows the rest wavelength of the line at the systemic
radial velocity of +25 km\,s$^{-1}$. Individual spectra are shifted by 0.15\,I$_{\rm c}$ with respect to each other.
Numerous telluric lines are clearly visible around the interstellar Na {\sc i} D--lines (5889 and 5895 \AA), especially
in the two bottom spectra. The sodium lines have stable positions and are shown here to  contrast the He line
variations and demonstrate the wavelength calibration accuracy. \label{f:Var}}
\end{figure}

Most absorption lines of the hot star are variable, especially
He {\sc i} lines which sometimes show emission components (see
Figure\,\ref{f:Var}). These components are seen in the line wings and
probably due to the fast rotating material in the inner parts of the
circum-primary disk. The detection of only weak emission supports
our determination of the primary's T$_{\rm eff}$, because He {\sc i}
lines transition into full emission typically occurs at T$_{\rm eff}
\sim 18000-20000$ K.

\begin{figure*}[t]
\setlength{\unitlength}{1mm}
\resizebox{15.cm}{!}{
\begin{picture}(150,100)(0,0)
\put (-5,50) {\includegraphics[width=7.3cm,bb = 0 0 311 258, clip=]{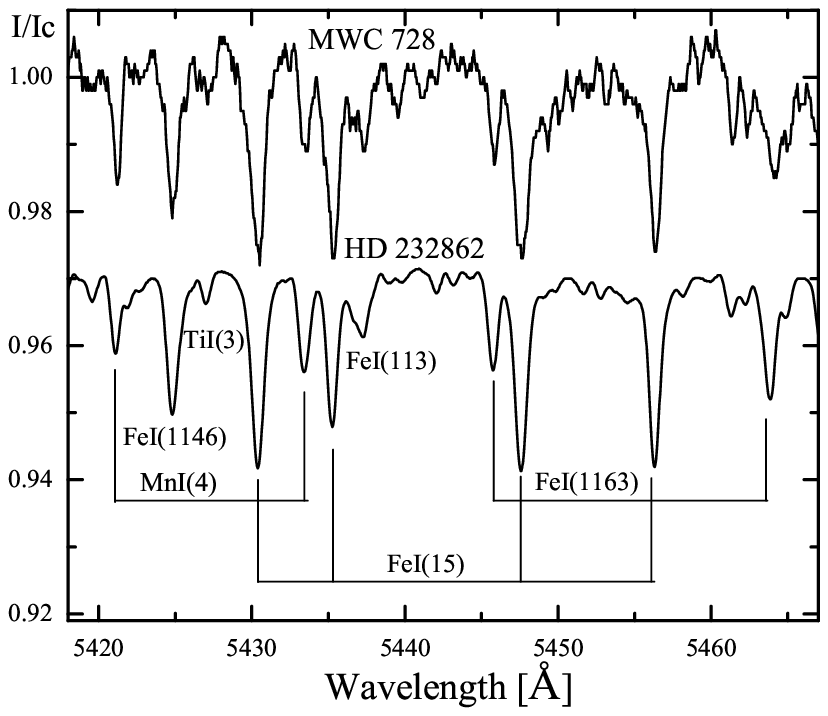}}
\put (55,50) {\includegraphics[width=7.3cm,bb = 0 0 311 258,  clip=]{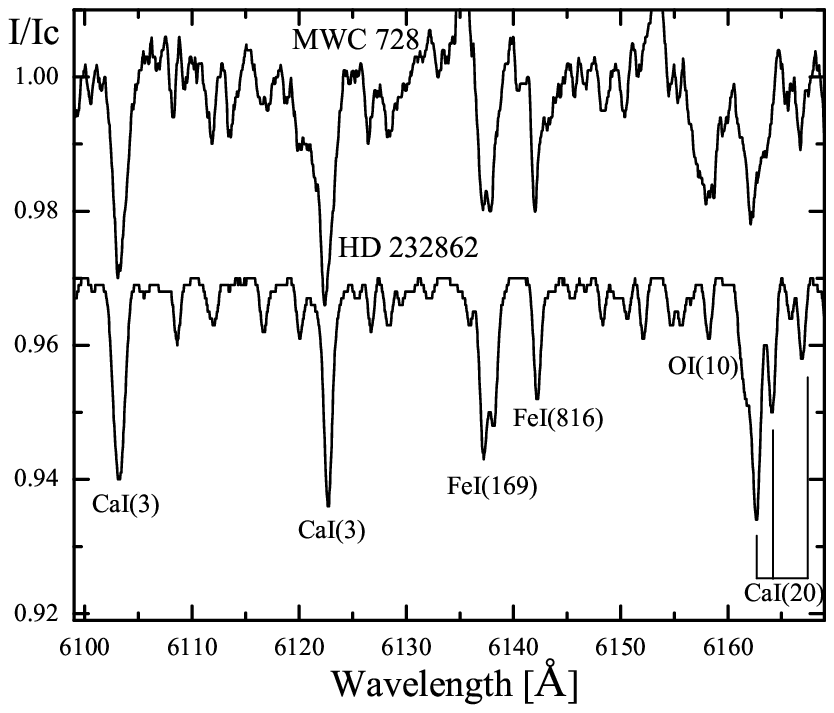}}
\put (115,50) {\includegraphics[width=7.3cm,bb = 0 0 311 258, clip=]{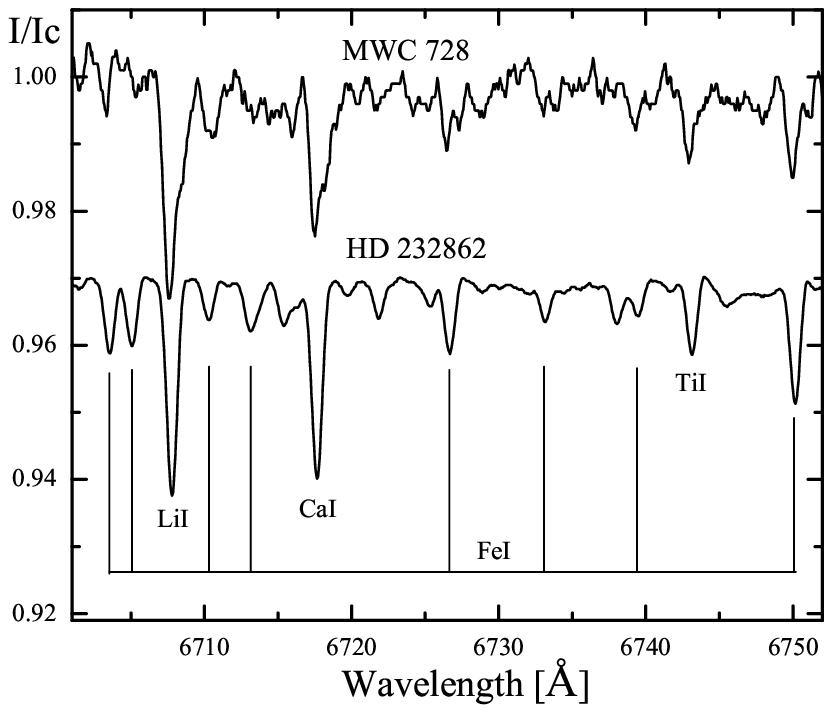}}
\put (-5,0){\includegraphics[width=7.3cm,bb = 0 0 311 258, clip=]{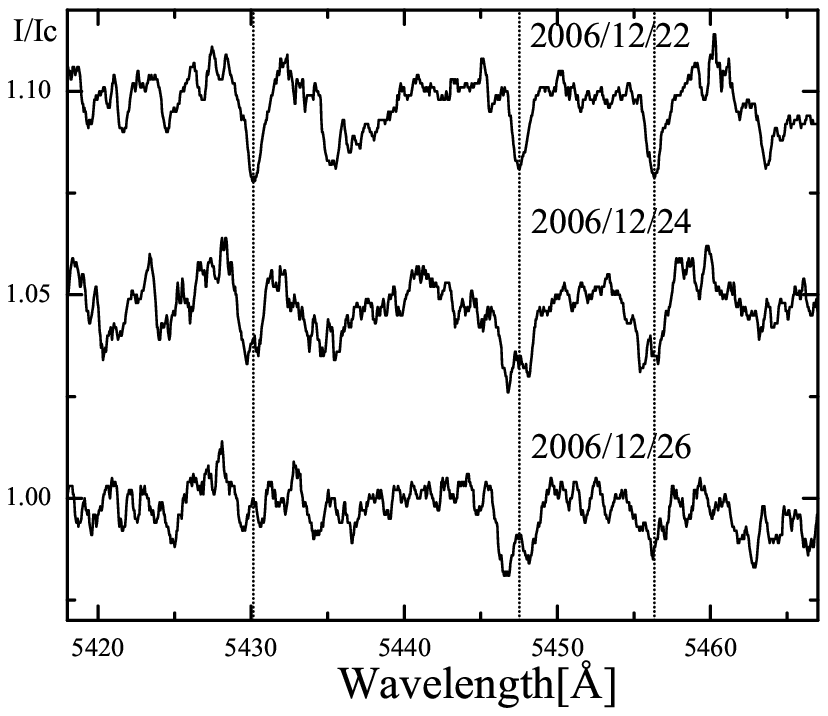}}
\put (55,0) {\includegraphics[width=7.3cm,bb = 0 0 311 258, clip=]{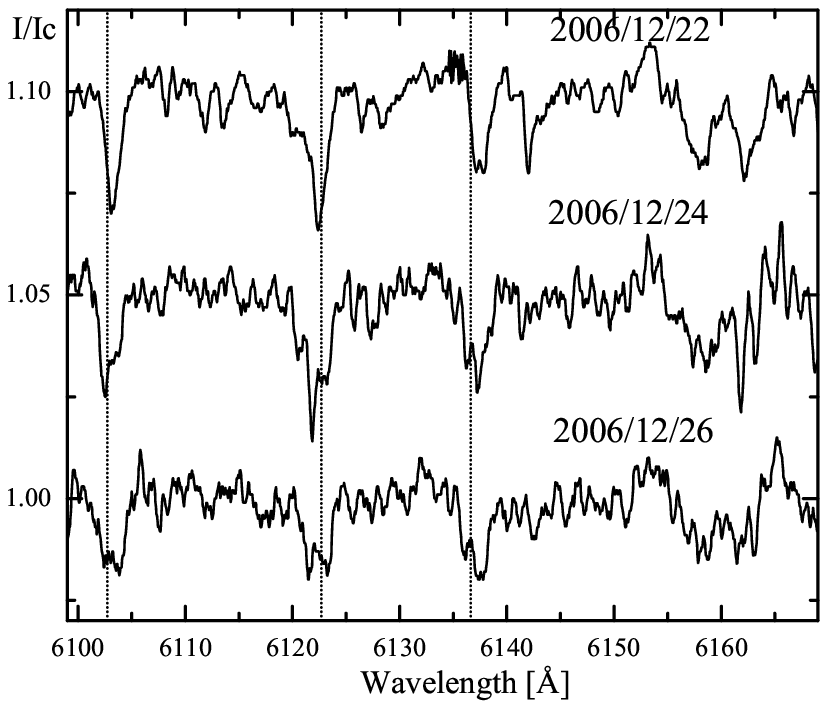}}
\put (115,0){\includegraphics[width=7.3cm,bb = 0 0 311 258, clip=]{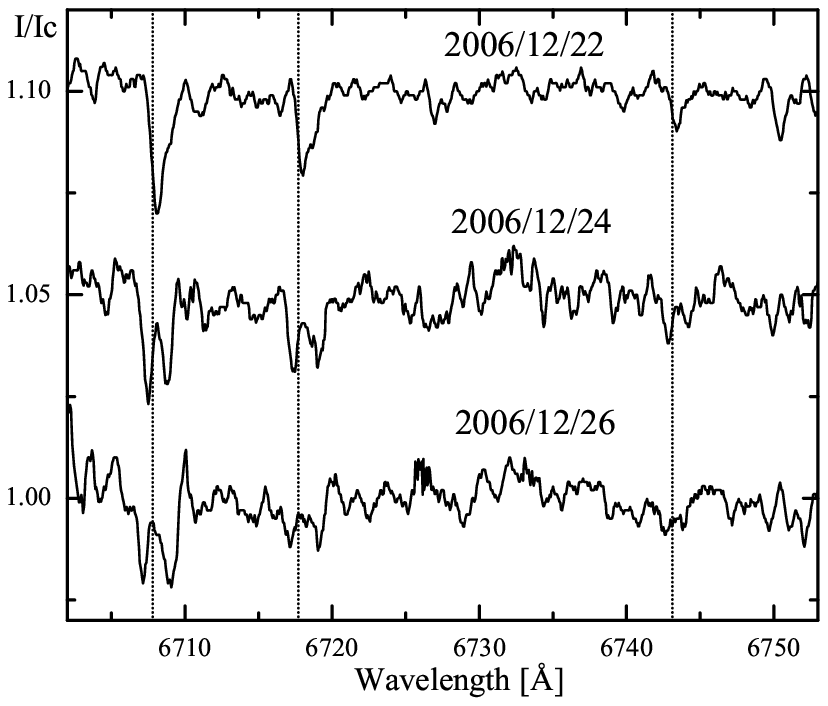}}
\end{picture}}
\caption{{\bf Top panels:} Absorption lines in the McDonald spectra
of MWC\,728 and HD\,232862 (G8 {\sc ii}, see text) for three spectral
regions. Lines of neutral metals are marked. {\bf Bottom panels:}
Examples of the short-term variability and splitting of some
absorption lines in different regions of the MWC\,728 spectra are
presented. Vertical lines show laboratory positions of the Ca {\sc
i}, Li {\sc i}, and Ti {\sc i}.
\label{f:AbsorptionLines}}
\end{figure*}

\begin{table*}[t]
\caption[]{Radial velocities of the absorption lines derived by
cross-correlation}\label{t4} {\small
\begin{center}
\begin{tabular}{ccrcccrcccrcccrc}
\hline\noalign{\smallskip}
JD  &  Phase&  RV &  Err. & JD  &  Phase&  RV &  Err.  & JD  &  Phase&  RV &  Err. & JD  &  Phase&  RV &  Err. \\
\noalign{\smallskip}\hline\noalign{\smallskip}
3363.730    &   0.000   &   21.9    &   1.1 & 4419.892    &   0.407   &   15.4    &   2.1 &   6244.761    &   0.769   &   27.1    &   1.1 &   6962.795    &   0.880   &   42.3    &   8.1 \\
3656.801    &   0.658   &   2.9     &   3.5 &  4421.886    &   0.480   &   10.0    &   3.4 &  6250.694    &   0.984   &   29.7    &   2.2 &  6963.764    &   0.915   &   41.0    &   3.7  \\
3721.654    &   0.016   &   20.7    &   1.9 & 4423.866    &   0.552   &   9.2     &   7.8 &    6295.634    &   0.619   &   4.0     &   2.1 & 6964.758    &   0.951   &   37.2    &   2.6\\
3726.711    &   0.200   &   22.6    &   1.9 & 4744.841    &   0.224   &   26.0    &   2.3 &   6320.678    &   0.529   &   19.2    &   1.7 &  6993.739    &   0.005   &   32.8    &   3.7\\
4081.875    &   0.115   &   24.4    &   3.1 & 4746.855    &   0.297   &   18.6    &   4.3 &   6322.655    &   0.601   &   15.7    &   2.7 &  6996.699    &   0.113   &   37.2    &   2.9\\
4082.658    &   0.144   &   38.1    &   3.2 & 4815.747    &   0.803   &   17.4    &   1.8 &   6580.981    &   0.995   &   40.4    &   2.2 &  7035.618    &   0.528   &   22.1    &   3.6\\
4083.884    &   0.188   &   33.3    &   3.7 & 5079.925    &   0.409   &   5.8     &   2.6 &    6584.811    &   0.135   &   38.5    &   2.8 & 7036.733    &   0.569   &   18.7    &   3.8\\
4092.612    &   0.506   &   16.0    &   3.5 & 5081.934    &   0.482   &   12.8    &   2.7 &   6585.786    &   0.170   &   39.7    &   2.3 &  7037.609    &   0.601   &   15.2    &   5.8\\
4094.576    &   0.577   &   2.5     &   3.1 &  5145.750    &   0.803   &   32.2    &   6.3 &  6587.831    &   0.244   &   32.8    &   2.3 &  7038.613    &   0.637   &   9.7     &   6.3\\
4096.750    &   0.656   &   21.3    &   4.7 & 5410.634    &   0.436   &   20.4    &   1.6 &   6628.678    &   0.730   &   14.0    &   3.5 &  7040.720    &   0.714   &   11.9    &   4.3\\
4412.847    &   0.151   &   34.6    &   5.0 & 5414.580    &   0.579   &   21.4    &   0.8 &   6961.798    &   0.844   &   37.9    &   1.9 &  7041.717    &   0.750   &   21.4    &   3.8\\
\noalign{\smallskip}\hline
\end{tabular}
\end{center}
}
\begin{list}{}
\item The list of  Julian dates  (2450000+) when spectra with signal-to-noise ratios
suitable for the absorption line measurements, column "Phase" contains
phases calculated for the 27.5-day orbital period, column "RV" lists
heliocentric radial velocities in km\,s$^{-1}$, and column "Err" shows
1--$\sigma$ uncertainties of the radial velocities in km\,s$^{-1}$.
\end{list}
\end{table*}

\begin{figure*}[t]
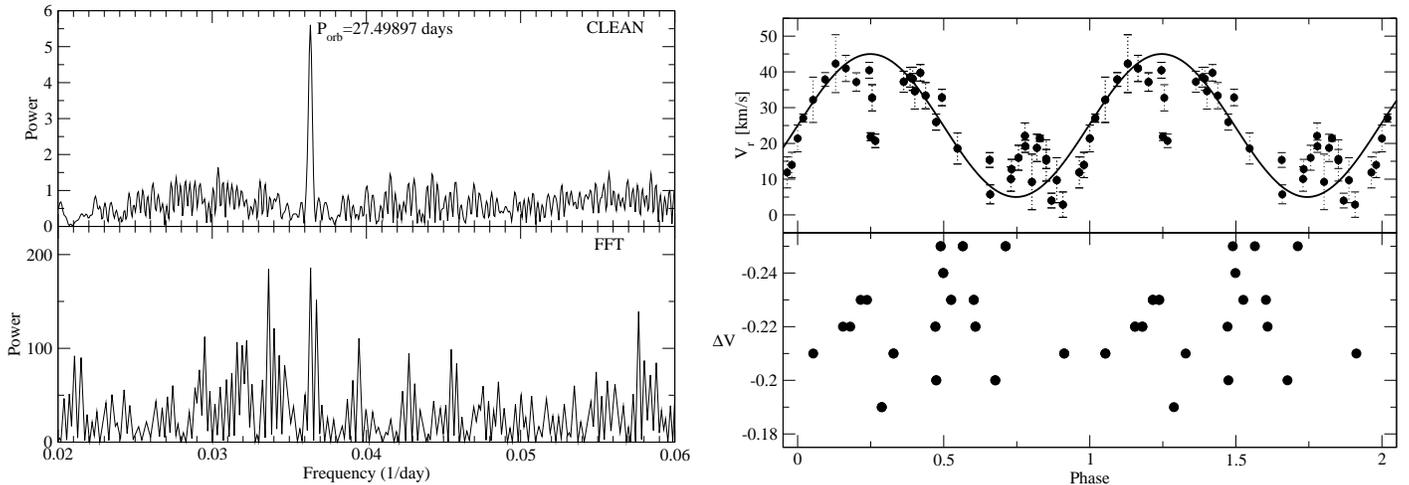

\setlength{\unitlength}{1mm}
\resizebox{15.cm}{!}{
\begin{picture}(150,65)(0,0)
\put (-5,0){\includegraphics[width=9.3cm, bb=40 50 725 530, clip=]{MWC728_fig7a.eps}}
\put (90,0){\includegraphics[width=9.25cm,bb=20 55 705 523, clip=]{MWC728_fig7b.eps}}
\end{picture}}
\caption{{\bf Left panels.} Power spectra
calculated by the DFT (bottom) and CLEAN (top) algorithms for
the absorption lines radial velocity derived by cross-correlation
(see text). {\bf Right panel.} Radial velocity (top) and the $V$--band photometry (bottom)
folded on the proposed orbital period P$_{\rm orb} $= 27.5 days. The $V$--band brightness is given
with respect to that of the comparison star, HD\,281193.
\label{f:Power}}
\end{figure*}

\begin{figure*}[t]
\begin{center}
\includegraphics[width=17.8cm,bb = 1 1 575 135, angle=0 ,clip=]{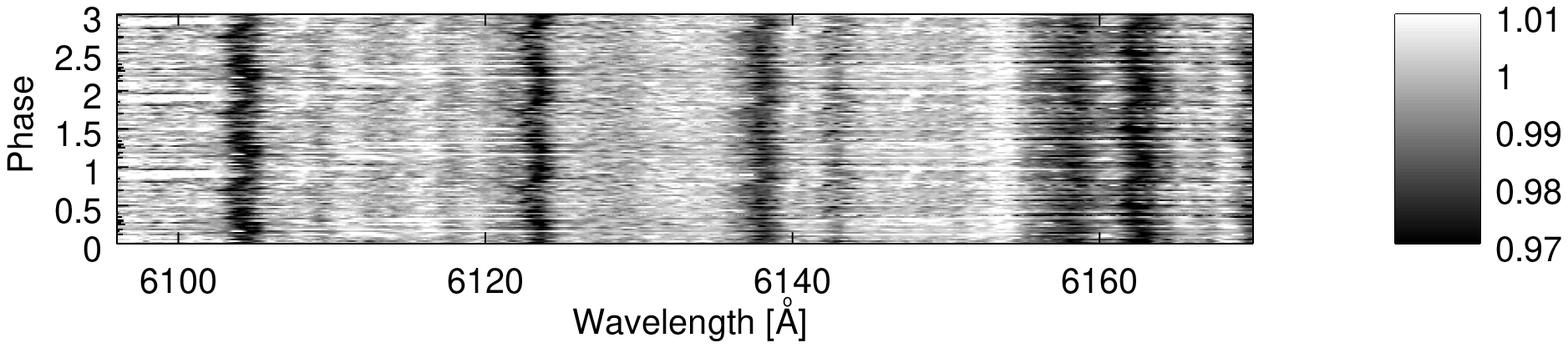}
\end{center}
\caption{Phased time series spectra in the 6100 -- 6170 \AA\ wavelength region folded three times (Y-axes)
on the orbital period of the system P$_{\rm orb} $= 27.5 days. The intensity scale in the continuum units is
shown in the right part of the graph.
\label{f:AbsorpTrail}}
\end{figure*}

The second group contains narrow lines of neutral metals, such as Fe
{\sc i}, Ca {\sc i}, O {\sc i}, Ti {\sc i}, and Li {\sc i} (see
Figure\,\ref{f:AbsorptionLines}). They become clearly detectable at $\lambda \sim$5300 \AA\
and can be traced all the way to the red end of our optical spectra
($\sim$10000~\AA). The weakness of these lines is caused by a stronger
contribution of the continuum from the hot star as shown below.
It is not easy to constrain the temperature of the cool companion in this
situation. The obvious absence of molecular features and the
presence of the Li {\sc i} 6708 \AA\ line, which is typically
observed in G5--K2 giants \citep[e.g.,][]{kumar11}, allows us to
assume a T$_{\rm eff} \sim 5000\pm1000$~K.

To put more constraints on the fundamental parameters of the cool
companion, we compared our spectra of MWC\,728 with those of Li-rich
giants obtained at the McDonald Observatory with the same instrumental
setup. The spectra were normalized to the continuum and smoothed with
the same box-car filter to get rid of a high-frequency noise.  A very good
match was found with the spectrum of the magnetic and Li-rich giant
HD\,232862 \citep[G8 {\sc ii}, T$_{\rm eff}$=5000K,][]{2009A&A...504.1011L}.
In the top panels of Figure\,\ref{f:AbsorptionLines} we compared the spectra of
MWC\,728 and HD\,232862 in selected intervals. To fit the absorption line
profiles of MWC\,728, we added a featureless continuum to the spectrum
of HD\,232862 in each interval. The best agreement was achieved with the
cool star contributions of 5\%, 7\%, 12\%, and 15\% in the 5410--5470 \AA,
6100--6170 \AA, 6705--6750 \AA, and 8320--8390~\AA\ intervals, respectively.

\begin{table*}[t]
\caption[]{Properties of the H$\alpha$ line in the spectra of MWC\,728}
\vspace*{-0.5cm}
\label{t5}
{\small
\begin{center}
\begin{tabular}{crrrcrcrrrcr}
\hline\noalign{\smallskip}
MJD           &I$_{\mathrm V}$ &  I$_{\mathrm R}$ &   EW      & Phase    &  Obs.  &  MJD&I$_{\mathrm V}$& I$_{\mathrm R}$& EW& Phase &  Obs.\\
\noalign{\smallskip}\hline\noalign{\smallskip}
3363.730    &   6.30    &   12.47   &   60.2    &   0.250   &   CFHT    &   6661.585    &   3.97    &   5.33    &   26.7    &   0.177   &   TCO \\
3656.801    &   3.70    &   6.61    &   39.5    &   0.908   &   SPM &   6670.291    &   5.49    &   5.88    &   31.0    &   0.493   &   Ondrj   \\
3721.654    &   5.85    &   6.22    &   37.2    &   0.266   &   McDon   &   6670.577    &   5.61    &   5.97    &   31.2    &   0.504   &   TCO \\
3726.711    &   5.63    &   8.38    &   44.9    &   0.450   &   McDon   &   6672.564    &   6.22    &   7.01    &   36.7    &   0.576   &   TCO \\
4081.875    &   6.20    &   8.06    &   50.2    &   0.365   &   SPM &   6674.564    &   5.84    &   7.13    &   37.7    &   0.649   &   TCO \\
4082.658    &   5.21    &   8.22    &   46.7    &   0.394   &   SPM &   6676.562    &   4.75    &   8.08    &   36.8    &   0.721   &   TCO \\
4083.884    &   5.62    &   8.18    &   42.8    &   0.438   &   SPM &   6677.588    &   4.52    &   8.72    &   37.9    &   0.759   &   TCO \\
4092.612    &   3.48    &   6.82    &   39.7    &   0.756   &   McDon   &   6678.553    &   3.92    &   7.58    &   32.1    &   0.794   &   TCO \\
4094.576    &   3.86    &   7.41    &   41.7    &   0.827   &   McDon   &   6680.543    &   4.23    &   7.58    &   33.7    &   0.866   &   TCO \\
4096.750    &   4.90    &   7.81    &   44.4    &   0.906   &   McDon   &   6682.525    &   4.86    &   7.73    &   34.3    &   0.938   &   TCO \\
4412.847    &   4.61    &   6.65    &   40.8    &   0.401   &   SPM &   6711.328    &   5.50    &   7.37    &   36.7    &   0.986   &   Ondrj   \\
4419.892    &   3.55    &   5.36    &   34.2    &   0.657   &   SPM &   6712.438    &   5.60    &   7.45    &   36.8    &   0.026   &   Ondrj   \\
4421.886    &   3.80    &   5.88    &   37.6    &   0.730   &   SPM &   6712.540    &   5.30    &   7.86    &   37.0    &   0.030   &   TCO \\
4423.866    &   4.45    &   7.77    &   47.9    &   0.802   &   SPM &   6717.273    &   6.06    &   6.15    &   35.9    &   0.202   &   Ondrj   \\
4744.841    &   4.71    &   6.67    &   34.3    &   0.474   &   SPM &   6861.564    &   4.24    &   9.22    &   43.0    &   0.449   &   Ondrj   \\
4746.855    &   6.00    &   6.49    &   33.0    &   0.547   &   SPM &   6948.702    &   4.75    &   8.64    &   38.7    &   0.618   &   TCO \\
4813.774    &   4.75    &   8.05    &   37.5    &   0.981   &   McDon   &   6949.725    &   4.76    &   8.41    &   37.8    &   0.655   &   TCO \\
4815.747    &   5.61    &   9.57    &   47.5    &   0.053   &   McDon   &   6955.744    &   5.09    &   7.49    &   32.6    &   0.874   &   TCO \\
5079.925    &   2.75    &   10.37   &   44.1    &   0.660   &   McDon   &   6956.777    &   5.26    &   7.42    &   35.5    &   0.911   &   TCO \\
5081.934    &   2.62    &   8.62    &   33.6    &   0.733   &   McDon   &   6957.727    &   4.89    &   6.69    &   33.7    &   0.946   &   TCO \\
5143.723    &   6.25    &   7.63    &   52.4    &   0.980   &   SPM &   6961.720    &   4.98    &   11.14   &   56.6    &   0.091   &   TCO \\
5145.750    &   4.71    &   8.82    &   50.2    &   0.053   &   SPM &   6961.798    &   5.28    &   12.05   &   61.9    &   0.094   &   SPM \\
5410.634    &   5.54    &   6.35    &   34.2    &   0.686   &   CFHT    &   6962.795    &   4.30    &   11.67   &   57.1    &   0.130   &   SPM \\
5414.580    &   5.64    &   6.35    &   34.1    &   0.829   &   CFHT    &   6963.764    &   3.67    &   11.16   &   50.3    &   0.166   &   SPM \\
5925.657    &   4.90    &   6.60    &   29.5    &   0.415   &   TCO &   6964.654    &   3.23    &   11.82   &   47.3    &   0.198   &   TCO \\
6154.576    &   2.84    &   7.05    &   29.2    &   0.739   &   Ondrj   &   6964.758    &   3.42    &   12.03   &   49.6    &   0.202   &   SPM \\
6158.531    &   2.57    &   5.91    &   25.7    &   0.883   &   Ondrj   &   6969.693    &   2.16    &   8.77    &   31.2    &   0.381   &   TCO \\
6187.475    &   4.40    &   5.56    &   27.8    &   0.936   &   Ondrj   &   6971.685    &   3.39    &   7.71    &   29.4    &   0.454   &   TCO \\
6244.761    &   5.16    &   7.89    &   44.1    &   0.019   &   SPM &   6972.701    &   3.80    &   7.18    &   28.8    &   0.491   &   TCO \\
6250.694    &   4.20    &   5.66    &   30.9    &   0.235   &   SPM &   6973.731    &   4.07    &   6.60    &   28.3    &   0.528   &   TCO \\
6290.675    &   4.02    &   7.42    &   23.3    &   0.689   &   McDon   &   6976.659    &   4.24    &   6.22    &   30.4    &   0.635   &   TCO \\
6295.634    &   6.13    &   5.71    &   38.3    &   0.869   &   McDon   &   6980.685    &   4.47    &   5.72    &   35.0    &   0.781   &   TCO \\
6320.678    &   3.45    &   5.58    &   22.1    &   0.780   &   SPM &   6982.656    &   4.32    &   6.43    &   33.9    &   0.853   &   TCO \\
6322.655    &   2.59    &   3.57    &   14.7    &   0.852   &   SPM &   6992.652    &   4.74    &   5.85    &   29.7    &   0.216   &   TCO \\
6580.981    &   4.83    &   4.54    &   33.7    &   0.246   &   SPM &   6993.718    &   4.62    &   6.61    &   36.5    &   0.255   &   SPM \\
6584.811    &   4.84    &   8.76    &   49.0    &   0.385   &   SPM &   6996.678    &   5.23    &   7.85    &   44.5    &   0.362   &   SPM \\
6585.786    &   4.45    &   9.19    &   49.3    &   0.420   &   SPM &   7002.668    &   2.89    &   6.42    &   24.1    &   0.580   &   TCO \\
6587.831    &   1.93    &   10.30   &   38.4    &   0.495   &   SPM &   7003.596    &   2.64    &   6.56    &   24.0    &   0.614   &   TCO \\
6589.851    &   1.97    &   8.92    &   28.2    &   0.568   &   SPM &   7004.645    &   2.95    &   6.30    &   23.8    &   0.652   &   TCO \\
6591.920    &   2.80    &   7.40    &   25.1    &   0.643   &   SPM &   7006.590    &   3.50    &   5.90    &   22.5    &   0.723   &   TCO \\
6592.607    &   2.49    &   6.63    &   21.4    &   0.668   &   Ondrj   &   7032.592    &   3.47    &   9.88    &   42.6    &   0.669   &   TCO \\
6592.956    &   2.97    &   7.89    &   24.8    &   0.681   &   SPM &   7033.570    &   2.62    &   10.28   &   38.8    &   0.704   &   TCO \\
6600.758    &   4.88    &   7.74    &   34.7    &   0.965   &   TCO &   7035.618    &   2.04    &   9.44    &   32.9    &   0.779   &   SPM \\
6605.735    &   4.26    &   7.95    &   38.3    &   0.146   &   TCO &   7036.733    &   2.56    &   9.02    &   31.6    &   0.819   &   SPM \\
6606.506    &   3.74    &   7.15    &   32.0    &   0.174   &   Ondrj   &   7037.609    &   2.44    &   8.39    &   29.2    &   0.851   &   SPM \\
6607.710    &   4.18    &   7.49    &   33.7    &   0.218   &   TCO &   7038.613    &   2.57    &   7.63    &   26.8    &   0.887   &   SPM \\
6611.709    &   4.88    &   6.49    &   30.4    &   0.363   &   TCO &   7039.559    &   3.28    &   7.29    &   25.6    &   0.922   &   TCO \\
6625.712    &   3.59    &   8.03    &   34.8    &   0.872   &   TCO &   7039.748    &   3.48    &   7.28    &   26.4    &   0.929   &   SPM \\
6626.663    &   4.20    &   7.21    &   35.0    &   0.907   &   TCO &   7040.720    &   3.43    &   6.35    &   26.5    &   0.964   &   SPM \\
6627.690    &   4.22    &   6.83    &   31.3    &   0.944   &   TCO &   7041.586    &   3.14    &   6.65    &   27.4    &   0.996   &   TCO \\
6628.678    &   4.33    &   6.52    &   30.1    &   0.980   &   TCO &   7041.717    &   3.25    &   6.51    &   27.4    &   0.000   &   SPM \\
6637.655    &   3.46    &   6.61    &   31.9    &   0.307   &   TCO &   7042.595    &   3.58    &   6.45    &   27.4    &   0.032   &   TCO \\
6639.658    &   3.67    &   7.62    &   36.6    &   0.379   &   TCO &   7044.508    &   4.24    &   5.89    &   25.9    &   0.102   &   TCO \\
6642.650    &   3.90    &   6.84    &   32.3    &   0.488   &   TCO &   7045.529    &   4.57    &   5.90    &   26.5    &   0.139   &   TCO \\
6643.691    &   3.81    &   7.18    &   33.4    &   0.526   &   TCO &   7059.535    &   5.96    &   5.71    &   36.5    &   0.648   &   TCO \\
6645.682    &   4.06    &   5.47    &   30.4    &   0.598   &   TCO &   7061.549    &   4.98    &   6.52    &   42.3    &   0.722   &   TCO \\
6653.641    &   7.27    &   7.69    &   52.0    &   0.888   &   TCO &   7062.544    &   5.68    &   8.14    &   47.2    &   0.758   &   TCO \\
6654.627    &   5.83    &   8.32    &   50.4    &   0.924   &   TCO &   7065.554    &   6.32    &   8.44    &   50.5    &   0.867   &   TCO \\
\noalign{\smallskip}\hline
\end{tabular}
\end{center}
}
\begin{list}{}
\item The columns lists Modified Julian dates (JD$-$245000), the blue (I$_{\rm V}$)
and red (I$_{\rm R}$) peak strengths in the continuum units, the
H$\alpha$ line EW in \AA, orbital phase according to the ephemeris
in the text, and the observing site ID (see Table\,\ref{t1}), respectively.
\end{list}
\end{table*}

\begin{figure}
\setlength{\unitlength}{1mm}
\resizebox{15.cm}{!}{
\begin{picture}(150,65)(0,0)
\put (5,0){\includegraphics[width=8cm,bb = 60 20 725 575, clip=]{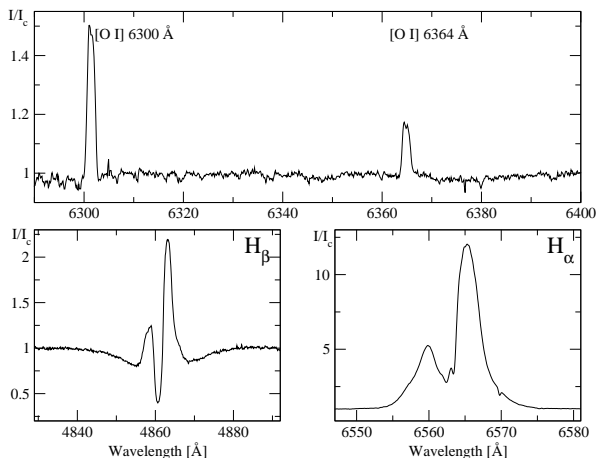}}
\end{picture}}
\caption{ Typical profiles of the H$\beta$, H$\alpha$ and [O {\sc
i}] emission lines in the spectrum of MWC 728. The spectrum was
taken on 10/30/2014 at OAN SPM. The wavelength scale is heliocentric, the intensity is
normalized to the nearby continuum. \label{f:OI}}
\end{figure}

The profiles and positions of the metallic absorption lines in the
spectra of MWC\,728 are noticeably variable (Figure\,\ref{f:AbsorptionLines},
bottom panels). Due to the line weakness that limits the accuracy of individual radial
velocity measurements, we used the cross-correlation method
implemented in the IRAF package {\it rvsao}. A wavelength region
between 6100~\AA\ and 6170 \AA\ that is well populated with the
absorption lines was chosen for the measurements. The region
includes the following lines: Ca {\sc i} (Mult.\,3) 6102.72,
6122.22, and 6162.17 \AA, Fe {\sc i} (Mult.\,169) 6136.62~\AA, Fe
{\sc i} (Mult.\,816) 6141.73 \AA, O {\sc i} (Mult.\,10) 6155.99,
6156.78, and 6158.19 \AA, and Ca {\sc i} (Mult.\,20) 6163.76 and
6169.56~\AA. No Li {\sc i} line at 6104.00 \AA\ can be clearly
separated from the Ca {\sc i} 6102.72 \AA\ line. Forty four spectra
obtained at CFHT, McDonald, OAN SPM, and TCO
were used to measure radial velocities.
A high signal-to-noise spectrum obtained at CFHT on 2004 December 24
was chosen as a template. The absorption lines are well defined in this spectrum
which allowed an accurate measurement of the averaged radial velocity of
21.9$\pm$1.1 km\,s$^{-1}$.
All radial velocities measured by cross-correlation are presented in Table \ref{t4}.

The Discrete Fourier Transform (DFT)
method\footnote{http://www.univie.ac.at/tops/Period04/}
\citep{1975Ap&SS..36..137D} and CLEAN procedure
\citep{1987AJ.....93..968R} were applied to search for periodicity
in the radial velocity data. The resulting periodograms are presented in
Figure~\ref{f:Power} (left panel). The two highest peaks in the DFT power
spectrum have nearly equal strengths at frequencies that correspond to
periods of 29.7 and 27.5 days, while the CLEAN procedure picks on the
only period of P$_{orb}$= 27.5(1) days. The latter one was adopted as an
orbital period of the system. The radial velocity curve and the
$V$--band photometric data folded on P$_{\rm orb}$ are shown in Figure
\ref{f:Power} (right panel). Figure\,\ref{f:AbsorpTrail} shows phased time series
spectra for the 6100--6170 \AA\ wavelength region where positional
variations of the absorption lines are clearly seen. The
ephemeris of the orbital phase from the radial velocity
measurements can be described as
$$ T_0 = HJD\ 2453356.855 + 27.499(108) \times E, $$
where $T_0$ corresponds to a moment when the cool companion is
located between the hot companion and the observer. The radial
velocity curve is symmetric about a $\gamma$--velocity of +25
km\,s$^{-1}$. It can be considered sinusoidal within the measurement
errors (circular orbit) and has a semi-amplitude of $\sim$20 km\,s$^{-1}$.
\begin{figure*}[!t]
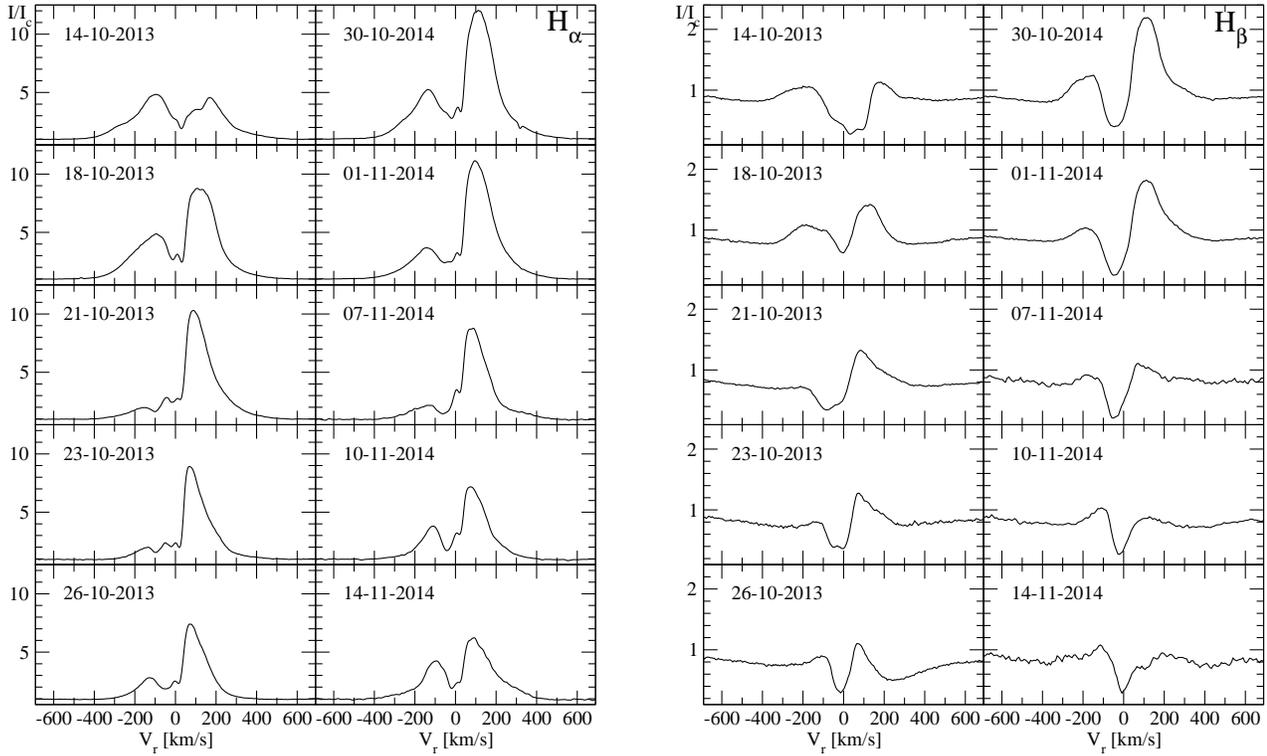

\setlength{\unitlength}{1mm}
\resizebox{15.cm}{!}{
\begin{picture}(150,110)(0,0)
\put (0,0){\includegraphics[width=8cm,bb = 66 48 585 707, clip=]{MWC728_fig10a.eps}}
\put (90,0){\includegraphics[width=8cm,bb = 66 48 585 707, clip=]{MWC728_fig10b.eps}}
\end{picture}}
\caption{ Short-term spectral variations of the H$\alpha$ (left
panels) and H$\beta$ (right panels) lines in the spectrum of
MWC\,728 in October 2013 and November 2014. The radial velocity
scale is heliocentric, the intensity is normalized to the nearby continuum. \label{f:Short-termHa}}
\end{figure*}

\begin{figure*}[t]
\setlength{\unitlength}{1mm}
\resizebox{15.cm}{!}{
\begin{picture}(150,70)(0,0)
\put (-5,0) {\includegraphics[width=10cm]{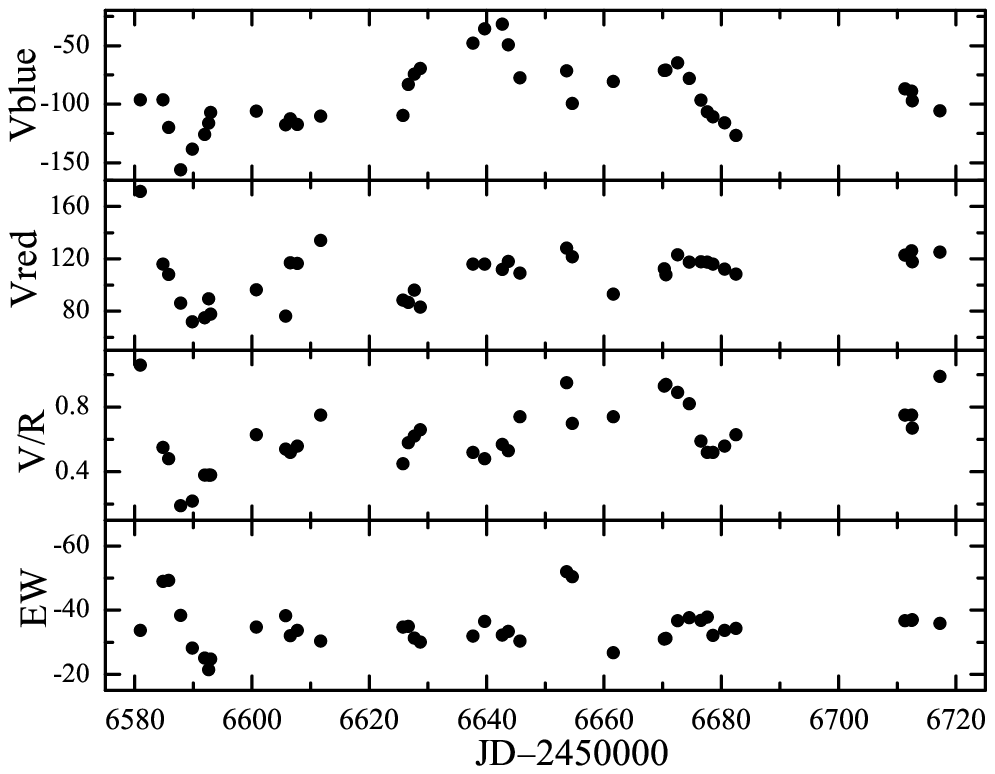}}
\put (90,0){\includegraphics[width=10cm]{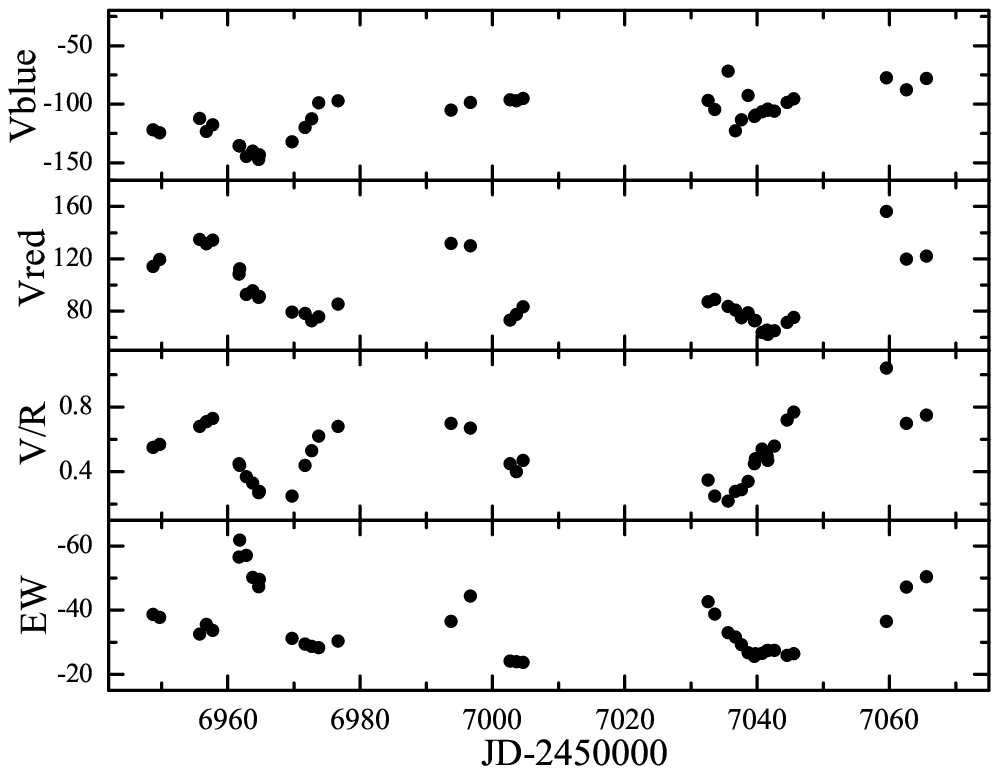}}
\end{picture}}
\caption{Variations of the H$\alpha$ line in the spectrum of
MWC\,728 detected during the 2013 (left panel) and 2014 (right
panel) observing seasons. From bottom to top are shown: equivalent
width (EW) in \AA, blue-to-red peak intensity ratio (V/R), heliocentric radial
velocity of the red (V$_{\rm red}$) and blue (V$_{\rm blue}$)
emission peak in km\,s$^{-1}$. \label{f:VarHa}}
\end{figure*}

\begin{figure}[!t]
\setlength{\unitlength}{1mm}
\resizebox{15.cm}{!}{
\begin{picture}(150,65)(0,0)
\put (0,0){\includegraphics[width=8.5cm,bb = 55 48 705 525, clip= ]{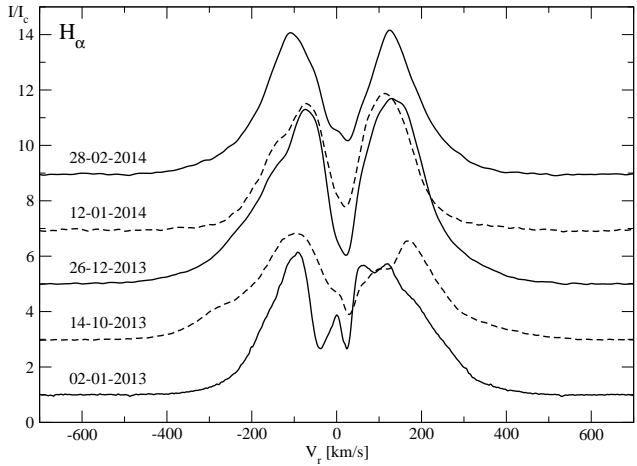}}
\end{picture}}
\caption{The profiles of the H$\alpha$ lines when V/R ratio is close to one.
The spectra are normalized to continuum and shifted along the Y--axis for the
best presentation. The radial velocity scale is heliocentric.The vertical dashed
line shows the systemic radial velocity of +25 km\,s$^{-1}$.
 \label{f:disk}}
\end{figure}

We note that sometimes the absorption lines
are split into two components. The splitting develops within a few
days and seems to occur at no preferential orbital phase. The
components appear on both sides of the initial unsplit line position
and are separated by up to 80 km\,s$^{-1}$. Such an episode observed
on 2006 December 22--26 is shown in Figure \ref{f:AbsorptionLines} (bottom panels).
Splitting occurs simultaneously in all absorption lines which adds
to the scatter of the radial velocities determined by cross
correlation. This phenomenon might be due to appearance of emission components
near the line centers, but its actual mechanism is unclear.

The third group of absorption lines has interstellar origin.
MWC\,728 is located at a galactic latitude of $\sim -20\arcdeg$,
nearly $3\arcdeg$ south of the association Per\,OB2.
The distance of Per\,OB2 from the Sun is
0.3 kpc \citep[e.g.,][]{md09}, and the interstellar extinction in
its line of sight is A$_V$ = 0.95 mag \citep{cernis93}. A
photometric study of a $\sim 1\arcdeg$ area around MWC\,728 by
\citet{r71} showed that the extinction in this direction is lower
and reaches $\sim$ 1 mag at a distance of $\sim$1 kpc. The
interstellar absorption lines in the MWC 728 spectrum are
represented by diffuse interstellar bands (DIBs), such as 4430,
5780, 5797, 6613~\AA, and the Na {\sc i} D--lines at 5889 and 5895
\AA. The sodium lines are narrow and saturated at our highest
resolution spectra ($R \sim$60000). Their measured average radial
velocity is 11$\pm$1 km\,s$^{-1}$. The DIBs are weak (e.g.,
EW$_{5780}$ = 0.13 \AA). Based on the relationship of the EW$_{5780}$
and E$(B-V)$ from \citet{herbig93}, the interstellar
reddening for MWC\,728 is E$(B-V) \sim$ 0.3 mag. This estimate
implies a visual interstellar extinction of A$_{\rm V} \sim$ 1.0
mag, and a distance to the object of at least 1 kpc. Both the
distance and reddening are discussed in more detail in
Sect.\,\ref{discussion}.

\subsection{Emission lines} \label{emission}

The emission-line spectrum is represented by the Balmer lines, the
[O {\sc i}] 6300 \AA\ and 6364 \AA\ lines (Figure~\ref{f:OI}), and
weak emission components of the He {\sc i} 5876 \AA\ line mentioned
above (Figure~\ref{f:Var}). The forbidden oxygen lines have
double-peaked profiles with nearly the same, slightly varying
intensity peaks and an average peak separation of 40~km\,s$^{-1}$.
The average radial velocity of these lines is $+27\pm5$ km\,s$^{-1}$
which virtually coincides with the $\gamma$-velocity of the absorption lines.
The equivalent widths of the [O {\sc i}] lines are
EW$_{6300} = -0.9\pm0.1$\AA \ and EW$_{6364} = -0.34\pm 0.05$\AA .

The H$\alpha$ and H$\beta$ lines are relatively strong and show
double-peaked emission profiles in most of our spectra
(Figure~\ref{f:OI} and \ref{f:Short-termHa}).
Measured parameters of the H$\alpha$ line are presented in
Table\,\ref{t5} and plotted in Figure\,\ref{f:VarHa} (for the seasons of 2013--2014). The H$\alpha$ line
profile is highly variable with changes occurring on a time scale
from days (Figure\,\ref{f:Short-termHa}) to months (Figure\,\ref{f:VarHa}).
The EW of the H$\alpha$ line emission component varies from $-$15~\AA\ to $-$61 \AA .

\begin{figure}[t]
\setlength{\unitlength}{1mm}
\resizebox{15.cm}{!}{
\begin{picture}(150,70)(0,0)
\put (0,0){\includegraphics[width=8.5cm,bb = 55 45 715 525, clip=]{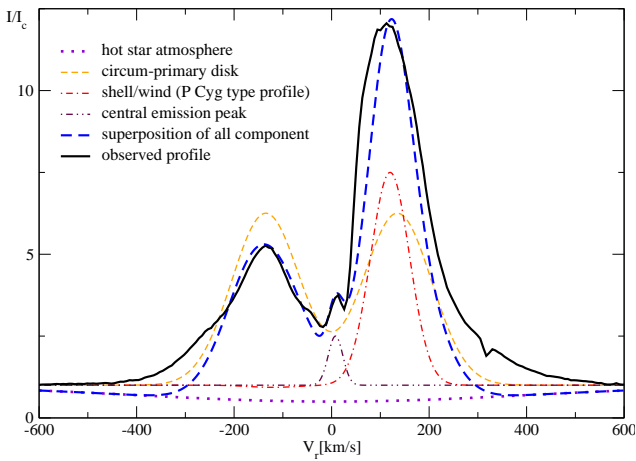}}
\end{picture}}
\caption{ A toy model of the H$\alpha$ profile of MWC\,728 observed
on 10/30/2014 at OAN SPM (see description in the text). The intensity and heliocentric radial velocity
are shown in the units as in Figure\,\ref{f:Short-termHa}. \label{f:ModelHa}}
\end{figure}

The blue peak of the H$\alpha$ line is usually weaker than the red one with
an average peak intensity ratio of V/R = 0.6$\pm$0.2.
This is different from
the Balmer line profiles of classical Be stars, in which the peaks typically show a
nearly equal strength (except for cyclic variations of the peak ratio, $V/R$
variations, that are due to density perturbations rotating in the disk).
This effect may be explained by emission from the stellar wind of the hot
companion that expands beyond its disk (see Figure\,\ref{f:ModelHa}).

\begin{figure*}[!th]
\begin{center}
\includegraphics[width=17.8cm,bb = 1 1 575 135, angle=0 ,clip=]{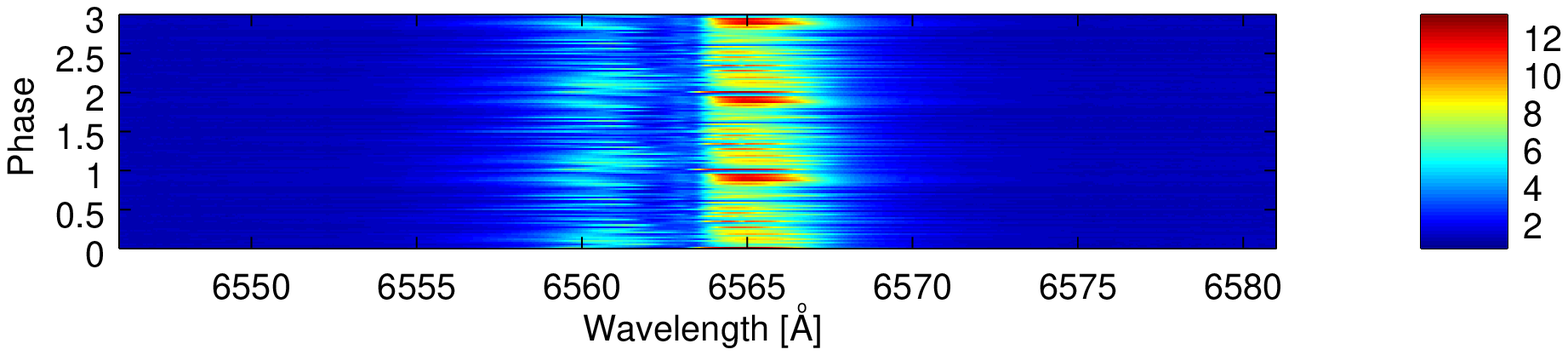}
\end{center}
\caption{Phased time series spectra around the H$\alpha$ line folded three times (Y--axis)
on the orbital period of the system  P$_{\rm orb} $= 27.5 day. The color coded intensity
scale in the continuum units is shown on the right.\label{f8}}
\end{figure*}

Sometimes the peak intensity ratio reaches V/R $\approx 1$
(Fig~\ref{f:disk}).
The H$\alpha$ line EW is typically $-35\pm3$ \AA\ at these
moments. However, on one occasion (2013 December 26) the line was
stronger with an EW of $-52$~\AA. We interpret the profiles with V/R
$\approx 1$ as dominated by the circum-primary disk emission with
almost no contribution from the primary's stellar wind.
The EW changes of the symmetric H$\alpha$ profiles
may be due to density variations in the circum-primary disk.

We do not find noticeable variations of the H$\alpha$ line profile
related to the periodicity of the absorption lines. However some
imprints of the orbital motion can be seen in the violet peak
in Figure~\ref{f8}, where phased time series spectra around
H$\alpha$ folded with the orbital period are shown.
The current amount of data is not sufficient to verify this suggestion.
A few months of continuous spectroscopic observations are needed
to explore the validity of this statement.

Another feature of the $H\alpha$ profile is a weak emission peak at a
nearly zero heliocentric radial velocity. It is clearly visible in most of
our high-resolution spectra but occasionally overlaps with one
of the main peaks (Figs.\,\ref{f:OI}, \ref{f:Short-termHa}, and~\ref{f:disk}).
The source of this emission is not clear yet.

To summarize our current understanding of the formation and
variations of the H$\alpha$ line we roughly reproduced it by a
simple toy model (Figure~\ref{f:ModelHa}). It includes an atmospheric
absorption from the hot star, a double-peaked emission from the
circum-primary disk, an emission from the shell/wind with a
P\,Gyg-type profile, and a weak emission peak at the systemic velocity.
All the components were represented by simple
Gaussian profiles. The resulting profile was calculated following
equation 12 from \citet{1985A&AS...62..339S} as $I_\lambda =
\Sigma\,I \times I_{P\,Cyg}^{abs}+ I_{P\,Cyg}^{em} $, where
$\Sigma\,I$ is a sum of all normalized fluxes of the underlying
components and $I_{P\,Cyg}^{abs}$, $I_{P\,Cyg}^{em} $ are absorption
and emission components from the P\,Cyg-type profile of the
shell/wind. Such a toy model can qualitatively reproduce
the observed profiles and their variability. However application of a
radiation transfer code that is capable of treating variable circum-primary
disk and wind components is required for quantitative modeling.
The latter is beyond the scope of this paper.

\subsection{Brightness variations}\label{brightness}

Photometric observations of MWC\,728 are sparse. It was observed in
the course of the Northern Stars Variability Survey
\citep[NSVS,][]{wozniak2004} in 1999--2000.
The object's brightness was detected to be $V$=10.17$\pm0.05$ mag
with no correction to the standard Johnson system. No regular
variations or evidence for eclipses were found in these data.

Our photometric data obtained at TShAO  show a scatter of $\sim$0.1
mag (see Figure\,\ref{f:Power}), similar to that of the NSVS data. However on two nights (2014
November 2 / JD2456964 and 2015 February 4--5 / JD2457058) we
managed to take photometry contemporaneously with times when the
emission-line spectrum got stronger. On both these occasions the
$EW_{H\alpha}$ increased by a factor of $\sim$1.5 from $\sim -40$
\AA\ to $\sim -60$ \AA, but the object faded by $\sim 0.1-0.25$ mag
in the $V$--band from the quiescence level. These variations of both
the spectral line and brightness can be explained by sporadic matter
outbursts from the surface of the hot star. Part of this
material obscures the stellar surface and leads to a fading, while
line emission becomes stronger due to increasing amount of
circumstellar gas. A similar phenomenon has been observed during the
built-up of the circum-primary disk around in the
$\delta$ Scorpii binary system \citep{2003A&A...408..305M}.

\section{Discussion}\label{discussion}

\subsection{The binary system parameters}\label{binary}

Adopting an orbital period of 27.5 days and a semi-amplitude of the
radial velocity curve of $K_{2} = 20$ km\,s$^{-1}$, we derive a mass
function of
 $$\frac{P_{\rm orb}K_2^3}{2 \pi G}=\frac{\rm{M}_{1}\, \sin^{3} {\it i}} {(1 + q)^2}\cong 2.3\times10^{-2}\rm{M_{\odot}},$$
where M$_{1}$ is the primary (hot) companion mass,
$q = \rm{M}_{2}/\rm{M}_{1}$ is the secondary to primary
mass ratio, and $i$ is the orbital inclination angle.

\begin{figure}
\setlength{\unitlength}{1mm}
\resizebox{15.cm}{!}{
\begin{picture}(150,60)(0,0)
\put (0,0) {\includegraphics[width=8cm]{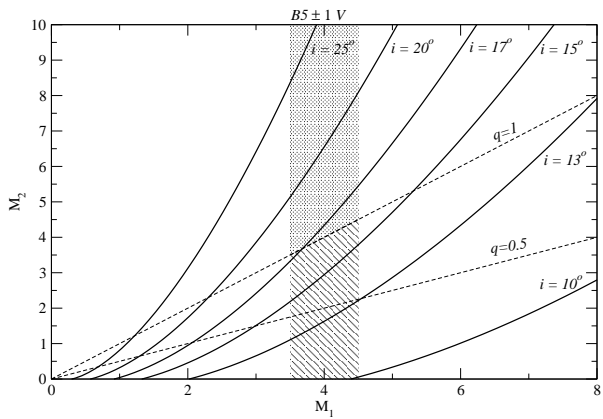}}
\end{picture}}
\caption{ M$_2$ vs M$_1$ relationship for different system inclinations.
The dashed lines correspond to mass ratios $q$ = 0.5 and $q$ = 1.
The shaded region marks the proposed mass range of the hot star based on
its spectral type. The part of this region below the $q$ = 1 line marks the possible range
of masses of the secondary.
\label{massfunc}\label{f:fm}}
\end{figure}

The small mass function implies a low inclination angle
to be reconciled with a meaningful hot star mass and the
absence of the photometric evidence for eclipses.
No signs of the orbital motion in the Balmer lines formed
in the circum-primary disk require a $q <  1$ (see Figure\,\ref{f:fm}).
Additionally, the detection of the radial velocity variations of the
absorption lines and the double-peaked  Balmer emission-line's
profiles are inconsistent with the system viewed exactly pole-on.

As mentioned in Sect.\,\ref{absorptions}, the He {\sc i} 4471~\AA\
and Mg {\sc ii} 4482~\AA\ line profiles suggest a $\it v \sin i \cong$ 110 km\,s$^{-1}$
for the hot star (Figure~\ref{f:HeMgTeff}, left panel).
Single main sequence stars with spectral types B5$\pm$1 have
masses of $3.5 M_\odot \lesssim M_1\lesssim4.5M_\odot$, radii of
$\sim$3.3 R$_{\odot}$, and breakup rotational velocities of
$\sim$420 km\,s$^{-1}$ \citep[e.g.,][]{Fremat05}. The presence of
a circum-primary disk suggests that the hot star rotates near the
breakup velocity, and an upper limit on the orbital inclination angle is
$\sim$15$\arcdeg$ if the circum-primary disk and the orbital plane
are coplanar. A more massive primary would require an $i < 10\arcdeg$
(Figure~\ref{f:fm}). Assuming $q = 1$ as an upper limit for the mass ratio
leads to the companions' separation of $\le74$R$_\odot$ and the size
of the secondary's Roche lobe of $\le$26R$_\odot$.

There is a potential problem in explanation of the Balmer line
profiles whose relatively deep central depressions (see, e.g.,
Figure\,\ref{f:HighBalmer} and \ref{f:OI}) may not be consistent with the derived low orbital
inclination angle. This apparent contradiction may be resolved by
assuming precession of the circum-primary disk. Such an explanation
has already been proposed to account for emission-line profile
variations in some Be stars \citep[e.g.,][]{Hirata07}. In the case
of MWC\,728, the precession period should be large as we did not detect
any noticeable changes in the emission-line profiles for over 10 years.

\subsection{Relative flux contributions}\label{relative_flux}

In order to place more tight constraints on the fundamental parameters
of the companions and the distance to the system we need to estimate relative
contributions of all the system constituents, i.e. both stars and the
circumstellar matter.  The cool companion contribution cannot be determined
precisely due to veiling by the other flux components that prevents us from
learning its true rotation rate and therefore the strength of its spectral lines.
Taking this uncertainty into account and using our results for the spectral line contributions
(see Sect.\,\ref{absorptions}), we adopt a $V$--band contribution of 10\%
from the cool companion to the total system flux.

Using stellar atmosphere models of \citet{Kurucz:1994aa}, we found that a
5000~K star continuum contributions to that of a 14000~K star
are nearly the same as the spectral line contributions in all the wavelength
intervals we mentioned in Sect.\,\ref{absorptions}). This implies that the disk
contribution to the total flux has nearly the same wavelength dependence as
that of the hot star at least between $\lambda \sim $5390~\AA\ and
$\lambda \sim$ 8800 \AA. A very similar result has been derived for optical
emission from disks of Be stars \citep[e.g.,][]{Carciofi2006} .

Since most of the disk surface seems to be facing the observer, it should
result in a non-zero disk contribution to the system radiation.
The $J$--band excess is mostly due to the free-free and bound-free disk
radiation, because circumstellar dust typically has temperatures below 1500 K
and dominates the spectral energy distribution at longer wavelengths (see
Figure\,\ref{f:spectrum}).
At the same time, we can estimate upper limits on the $J$--band excess and
interstellar extinction by assuming no circumstellar contribution to the continuum
in the optical range. Using our average photometric data for the dates when no
strong spectral variations were observed ($V = 9.77$ mag, $B-V = 0.27$ mag, see
Table\,\ref{t2}) and taking into account the contribution from the cool companion, we
get the following upper limits: $E(B-V)$ = 0.39 mag, $A_V$ = 1.22 mag, and the $J$--band
excess of $\sim$0.3 mag.

Based on these results, we adopt a lower interstellar extinction
of $A_V \sim$ 1.0 mag, which is consistent with that derived from the diffuse
interstellar band EW (see Sect.\,\ref{absorptions}) and implies a
circumstellar color-excess of $E(B-V) \sim$0.1 mag. This estimate does not
contradict our observations of the He {\sc i} 4471 \AA\ and Mg {\sc ii} 4482 \AA\
lines and suggests a disk contribution of $\le$20 \% to the continuum at these wavelengths.

Finally, we estimate the expected contributions of
the hot companion, the disk, and the cool companion in the optical
range (in the $V$--band) to be roughly 60\%, 30\%, and 10\%,
respectively. This is in agreement with a general conclusion that
the disk contribution to the total optical flux in Be stars may
hardly exceed 50\% of the hot companion flux
\citet{2012ApJ...756..156H}. The wavelength dependence of the disk radiation
in the IR region was calculated by extrapolation of the difference between the
total flux and that of both stars (see Fig\,\ref{f:spectrum}).

The fractional contributions define a lower limit
the secondary's radius $R_2 = 7.7R_\odot$ \citep[corresponding to a
luminosity type {\sc iii},][]{Straizys1981}, if we
assume a typical main-sequence mass of $M_1=4.0 M_\odot$ and a
radius of $R_1=3.3R_\odot$ for the primary. The companions would have these radii at a distance
of 1 kpc, which is consistent with the distance dependence of the
interstellar extinction (see Sect.\ref{absorptions}) and
can be considered a lower limit of the system distance. An upper limit is $\sim$3.5\,kpc
at which the secondary fills its Roche lobe.

\begin{figure}
\setlength{\unitlength}{1mm}
\resizebox{15.cm}{!}{
\begin{picture}(150,75)(0,0)
\put (-2,0) {\includegraphics[width=9cm,bb= 15 18 270 230,clip=]{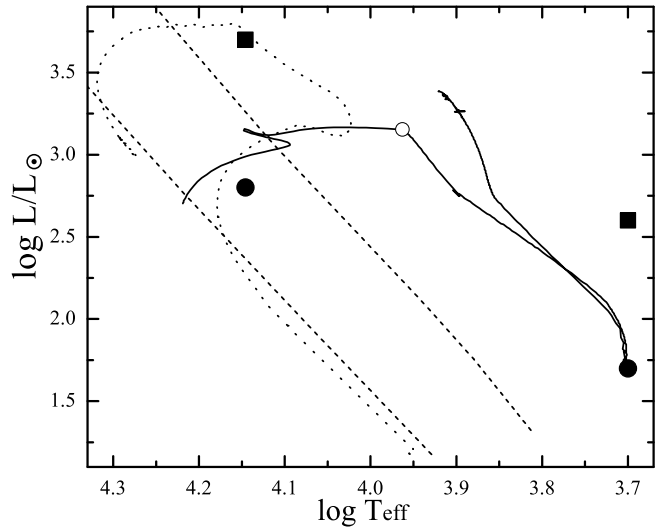}}
\end{picture}}
\caption{{Evolutionary tracks of a 5 M$_{\odot}$ (solid line) and
a 2 M$_{\odot}$ (dotted line) components of a close binary system with
non-conservative mass loss from \citet{vanRensbergen08}. The onset
of mass transfer is marked with an open circle on the track of the 5
M$_{\odot}$ star, while the 2 M$_{\odot}$ star is still
near the starting point of its evolution. The tracks are shown until
the end of the mass transfer stage. The dashed lines show the zero-age and terminal-age
main sequence for rotating single star models \citep{Ekstrom:2012aa}. The filled circles
show the fundamental parameters of both companions for the lower limit on the
system distance, and the filled squares show them for the higher distance limit (see Table\,\ref{t6}).}
\label{f12}}
\end{figure}

Fundamental parameters of the companions for both distances are
listed in Table \ref{t6}. The upper distance limit is hardly
consistent with the orbital parameters, because it corresponds to a
low critical rotational velocity of the hot star ($\sim$320
km\,s$^{-1}$). The latter requires $i \ge 20^{\circ}$ and a cool
companion that is more massive than the hot star (see Figure\,\ref{f:fm}).
Therefore, it is unlikely that the cool companion fills
its Roche lobe. This allows us to assume that the distance to the
system is close to its lower limit, and the companions mass ratio is $q \sim$0.5.

\begin{table}
\caption[]{Fundamental parameters of the stellar companions of the
MWC\,728 binary system at different distances}\label{t6}
\begin{center}
\begin{tabular}{ccrcr}
\hline\noalign{\smallskip}
D, kpc & $\log L_1/L_{\odot}$ & $R_1/R_{\odot}$ & $\log L_2/L_{\odot}$ & $R_2/R_{\odot}$ \\
\noalign{\smallskip}\hline\noalign{\smallskip}
1.0    & 2.8 &  3.3   & 1.6 &  7.7 \\
3.5    & 3.7 & 15.0   & 2.6 & 26.0 \\
\noalign{\smallskip}\hline
\end{tabular}
\end{center}
\begin{list}{}
\item Column 1 lists the distance, columns 2 and 3 show the
luminosity and radius of the hot star, columns 4 and 5 show the
luminosity and radius of the cool star.
We assumed T$_{\rm eff}$ = 14000 K and BC$_V = -1.1$ mag for the hot
star, T$_{\rm eff}$ = 5000 K and BC$_V = -0.3$ mag for the cool
star. The bolometric corrections are taken from \citet{m97}.
\end{list}
\end{table}

\subsection{Possible system model}\label{sysmodel}

The strong IR excess at $\lambda \ge 2 \mu$m (see
Figure\,\ref{f:spectrum}) implies the presence of dust in the system.
It should exist in the circum-binary area, as the stars are too
close together to allow its existence around either star.
\citet{m07} proposed that the dust forms out of the material lost
from the system in the course of mass-transfer between the
stars. Evolution of interme\-diate-mass binary systems has been
studied theoretically by \citet{vanRensbergen08}.
They calculated a grid of models for conservative evolution
as well as for a weak and strong tidal interaction between the stars.
The grid includes detailed calculations of physical and orbital parameters
from formation through the entire mass-transfer phase and beyond.
Non-conservative (liberal) evolution, when some mass is
lost from the system, was only considered for the more massive
companion mass of $\ge$5 M$_{\odot}$.

Browsing through the grid, we found a model which qualitatively
explains our findings for MWC\,728. It has initial
masses of 5 M$_{\odot}$ and 2 M$_{\odot}$ and an
initial orbital period of 8 days. The mass-transfer starts
in $1.12\,10^{8}$ years since the evolution began when the more
massive star fills its Roche lobe. It lasts for $\sim 5.5\,10^{5}$
years when the donor reaches a mass of 0.9 M$_{\odot}$, the gainer
has 4.2 M$_{\odot}$, and nearly 2 M$_{\odot}$ is lost from the
system. Therefore there is enough material to produce dust in the
circum-binary area. The evolutionary tracks for the stars in this
model are shown in Figure\,\ref{f12}.

Although the orbital period evolution in the model does
not match that of MWC\,728 (it reaches
27 days in the model, but only when the companions have almost the same
temperature and luminosity), it is qualitatively consistent with the
nature of this system. In parti\-cular, the companions' luminosity and
their ratio along with their temperatures roughly match our results.
Refining the system parameters with further observations and
calculating a finer grid of binary models with non-conservative
evolution would provide a more complete explanation of
the system properties.

\section{Conclusions}\label{conclusions}

Our study confirm an earlier suggestion by \citet{m07a}
that MWC\,728 is a binary system. We found the orbit to be circular (within
the measurement errors)  with a period of P$_{\rm orb}$ = 27.5 days. The
system consists of a B5 primary that is $\sim$6 times brighter (in the $V$--band)
than a G8 {\sc iii} secondary and is most likely located at a distance of $\sim$1 kpc.

The presented results further support the idea that FS\,CMa
objects are a group of moderately evolved binary systems and not
proto-planetary nebulae or young stars. Although nearly a dozen
FS\,CMa objects shows signs of binarity
mostly through the presence of spectral lines of a cool star
\citep{m07,m07a,m11}, MWC\,728 is only the second group object with
detected orbital motion. The first one was CI\,Cam \citep[P$_{\rm
orb}$ = 19.41 days,][]{Barsukova2006}, and it seems to have a
degenerate secondary companion. In all recognized FS\,CMa type binaries
but one \citep[MWC\,623,][]{Zickgraf2001} spectral lines of the secondary
companion are very weak and require both high-resolution and high
signal-to-noise spectroscopy to detect them. Additionally, irregular
variations of the emission-line spectrum contribute to veiling of
the hot star orbital motion. This phenomenon occurs on a time scale of
a few days and may represent fast variations of the hot star wind specific
to this evolutionary stage. These findings call for frequent and long-term
coordinated spectroscopic and photometric campaigns to reveal the
objects' nature.

A recent result by \citet{de-la-Fuente:2015aa} on two newly found FS\,CMa
group candidates in young clusters brought up an idea that some
FS\,CMa objects could result from a merger in a binary system. This
is definitely a possibility, although failure to detect of signs of the
secondary companion does not unequivocally imply its absence.
Models of non-conservative binary evolution (see
Sect.\,\ref{discussion}) shows that the cooler secondary may be much
fainter than the hot primary for a long time. In any case, the
binary model remains the major mechanism to explain
objects of this group.

We will continue coordinated spectroscopic and photometric
observations of MWC\,728 in order to further study properties of
occasional increases of the emission-line spectrum. They seem to be
associated with the object's brightness decreases and should help to
better constrain the nature of the binary system.

\acknowledgements A.~M. acknowledges financial support from the
University of North Carolina at Greensboro and from its Department
of Physics and Astronomy. A.~M.,  S.~Z. and D. K. acknowledge support from
DGAPA/PAPIIT project IN100614 and CONACyT grant CAR 208512.
We also acknowledge support from the programs
MSM0021620860 of the Czech Ministry of Education and
F.0679 ``Astrophysical studies of stellar and planetary systems'' of
the Science committee of the Ministry of education and science of
the Republic of Kazakhstan. We thank M.~Krugov, R.~Kokumbaeva, and
I.~Reva for obtaining and reducing photometric data at the Tien-Shan
Observatory. This research has made use of the SIMBAD
database, operated at CDS, Strasbourg, France.

\bibliography{ms}{}

\end{document}